\documentclass[final]{agujournal2019}
\usepackage{url} 
\usepackage[inline]{trackchanges} 
\usepackage{soul}
\draftfalse

\journalname{AGU Advances}

\begin{document}

%
%


\title{Neural Operators Reveal Scale-Dependent Limits on Recovering Greenland Ice Loss from Sea Level Fingerprints}

%
%




\authors{
Xiyuan Bao\affil{1},
Sophie Coulson\affil{2},
Natasha Valencic\affil{1},
S. Mostafa Mousavi\affil{1},
S\"{o}nke Dangendorf\affil{3},
Andrew Jason Lloyd\affil{4},
and Jerry X. Mitrovica\affil{1}
}

\affiliation{1}{
Department of Earth and Planetary Sciences,
Harvard University,
Cambridge, MA 02138, USA
}

\affiliation{2}{
Department of the Geophysical Sciences,
The University of Chicago,
Chicago, IL 60637, USA
}

\affiliation{3}{
Department of River-Coastal Science and Engineering,
Tulane University,
New Orleans, LA 70118, USA
}

\affiliation{4}{
Lamont-Doherty Earth Observatory,
Columbia Climate School,
Columbia University,
Palisades, NY 10964, USA
}




\correspondingauthor{Xiyuan Bao}{xiyuanbao@fas.harvard.edu}

\begin{keypoints}
\item Greenland sea level fingerprints constrain ice loss with scale dependent and prior dependent resolving power
\item Weak higher order fingerprint principal components retain fine scale ice information despite explaining little sea surface height variance
\item Relying purely on the fingerprint to recover Greenland's ice loss hits an accuracy ceiling, capping its ability to explain observed changes at $R^2\simeq0.79$.
\end{keypoints}

\begin{abstract}
Sea level fingerprints connect the spatial distribution of rapid ice mass change to sea surface height (SSH) change, but the spatial resolving power of this inverse problem remains poorly quantified. Here we provide the first systematic, scale-resolved quantification of Greenland Ice Sheet mass change structure recoverable from its SSH fingerprint when ice loss geometry varies. We develop a physics-informed neural operator, constrained by an accurate neural-operator emulator of the forward sea level equation, to probe the information content of this ill-conditioned inverse problem. Across synthetic ice change ensembles, broad ice loss patterns can be recovered, but distinct ice fields can produce nearly identical fingerprints, so fine-scale structure depends on the assumed \textit{a-priori} ice change distribution. Inversion fidelity varies with the wavelength and amplitude of ice mass change. An empirical error law separates these effects and defines a scale-dependent accuracy ceiling. For a satellite-altimetry-derived reference field, the estimated fingerprint-only ceiling is $R^2\sim0.79$ under the present SSH observation geometry and matched-prior assumption. The leading three principal components explain 99.76\% of fingerprint variance, yet higher-order components retain information needed to distinguish finer ice change structure. SSH observations near Greenland carry most information on ice loss locations. Applying the leading fingerprint principal components to the observed 1993--2019 SSH rate recovers the large-scale Greenland fingerprint, while the residual field is correlated with basin-scale bathymetry, consistent with unresolved sterodynamic variability. These results establish a quantitative empirical bound on fingerprint-only ice-loss resolving power and identify how independent observations and ice sheet constraints can improve inference.
\end{abstract}
\section*{Plain Language Summary}

As Greenland loses ice, sea level does not rise uniformly around the world. Changes in gravity, Earth deformation, and rotation produce a characteristic geographic pattern of sea level change, known as a sea level fingerprint. We explore how much information about the location, spatial scale, and magnitude of Greenland ice loss can be recovered from this fingerprint alone. Using Artificial Intelligence models trained on numerical sea level calculations, we find that while broad patterns of ice loss can be recovered, increasingly fine-scale changes become unconstrained and different ice loss patterns can produce the same (within uncertainty) sea level response. Some information about these finer changes remains in weak spatial features of the fingerprint, but these features are also more difficult to separate from ocean variability in real observations. We also find that sea level observations within a few hundred kilometers of Greenland contain much of the information useful for locating ice loss. These results quantify what sea level observations alone can reveal about ice sheet mass change and show where improved observations and additional ice sheet constraints can most effectively increase that information.

\section{Introduction}

Rapid mass loss from land ice changes sea level globally, but the response is spatially nonuniform because the redistribution of ice and ocean mass perturbs the gravitational field, deforms the solid Earth, and changes Earth rotation. The resulting spatial patterns, commonly referred to as sea level fingerprints, connect the geometry of land-ice mass change to a corresponding pattern of sea-surface-height (SSH) change \cite{farrell1976post,mitrovica2011robustness}. The geometry of ice loss is best constrained using near field sea level observations: ice loss distributions with the same total mass change can produce very different regional sea level responses, but progressively convergent responses farther from the ice sheet \cite{mitrovica2011robustness,cederberg2023consistency}. The geographic variability of the sea level response therefore contains information about both the magnitude and spatial distribution of ice mass change. We focus on the SSH change that can be compared directly with the altimetry observations, instead of the full relative sea level, and we use the term SSH fingerprints interchangeably with sea level fingerprints throughout the text.

Spatial sea level observations have long been used to constrain the distribution of ice mass change. \citeA{mitrovica2001recent} showed that geographically distinct sea level patterns evident in tide-gauge records could be used to estimate contributions from different polar ice reservoirs. Sea level fingerprints have since been used to quantify the regional and global sea level effects of contemporary ice and water mass redistribution \cite{riva2010fingerprint,bamber2010fingerprint,hay2015probabilistic,hay2017robustness,dangendorf2024probabilistic}, to construct projections of future sea level change \cite{slangen2014projecting}, and to separate mass and sterodynamic contributions in regional and global sea level budgets using a variety of observations \cite{rietbroek2016revisiting,dangendorf2021data}.  \citeA{coulson2022detection} detected the fingerprint of contemporary Greenland Ice Sheet (GrIS) mass loss in satellite-altimetry-derived SSH rates from 1993-2019. Building on this detection, \citeA{coulson2026inversion} performed the first direct inversion of the observed SSH rates for GrIS mass change using a linear Bayesian framework and demonstrated strong non-uniqueness in the inference: high-dimensional ice mass change over the entire GrIS could fit the SSH observations while producing unrealistic fields of ice mass change, whereas limiting ice loss to a suite of predefined, spatially fixed disks yielded robust estimates in different drainage basins. These results raise a more general question of what spatial scales of ice mass change are resolved by the fingerprint observations even in the case of error free observations.

This question is, of course, within the class of problems related to determining the resolving power of a geophysical inverse problem \cite{backus1968resolving,backus1970uniqueness}. The sea level response spatially integrates the effects of ice loading, so short-wavelength variations in ice mass change produce progressively weaker changes in the SSH field and different ice fields can generate similar fingerprints \cite{crawford2018quantifying}. Classical inversions address this loss of information through regularization or prior constraints, for example by penalizing roughness, restricting the solution basis, or prescribing a covariance structure \cite{tikhonov1963solution,parker1994geophysical}. Such constraints stabilize poorly resolved directions of the inverse problem, while the resulting spatial resolution depends on both the observations and the imposed prior. In the case of the Greenland SSH fingerprint, the resolving power depends on the spatial scale and magnitude of the ice mass change scale, \textit{a-priori} assumptions of the mass loss geometry, and where the corresponding ice-loss information resides within the smooth SSH field. These issues, and thus the associated resolving power, remain unclear.

Addressing these questions requires repeatedly evaluating the forward and inverse mappings over large ensembles of spatially distributed ice change fields without reducing the ice loss pattern to a few prescribed parameters. Neural operators are well suited to this setting because they learn field-to-field mappings and, once trained, can be evaluated efficiently across many spatial realizations \cite{li2021fourier}. Here we use controlled ensembles spanning different ice change scales, amplitudes, and spatial priors, enabled by a physics-informed neural-operator framework. We use these ensembles to quantify the resolving power of the Greenland fingerprint, determine how information about the spatial scale of ice change is distributed within the SSH field, and examine how much of the resulting fingerprint structure can be identified in observed SSH rates. Importantly, this framework allows the information supplied by the fingerprint itself to be separated from information introduced through the ice mass change prior.

\begin{figure*}
  \centerline{\includegraphics[width=\linewidth]{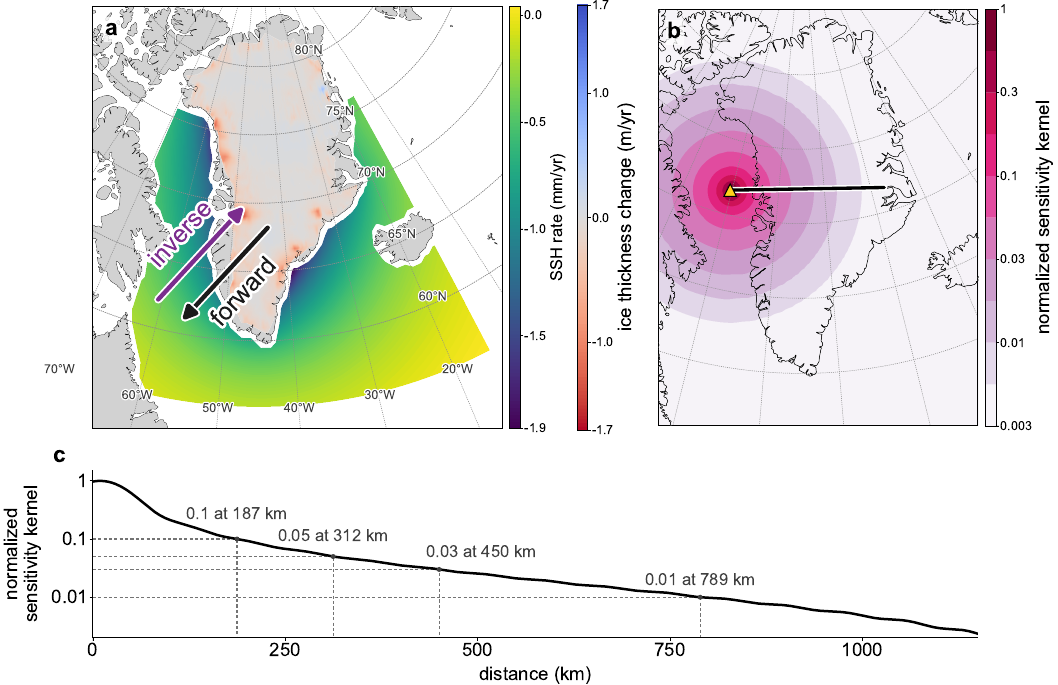}}
  \caption{Setup of the sea level fingerprint problem in this study. (\textit{a}) The forward problem (the black arrow) solves for the sea-surface-height (SSH) rate given ice-thickness change across the Greenland ice sheet, while the inverse problem is in the opposite direction (the purple arrow). The reference case is shown, with the fingerprint of the satellite-altimetry-derived Greenland ice loss \cite{simonsen2021greenland}, on a Gauss-Legendre grid of spherical harmonic degree and order 512. (\textit{b}) The normalized SSH sensitivity kernel to ice change at each location is shown for an example SSH site at 71.63 $^\circ$N, 60.37 $^\circ$W marked by the triangle (\textit{c}) The same kernel along the eastward great-circle transect toward the ice sheet interior.}
\label{fig:setup}
\end{figure*}

\section{Methods}
In short, our approach learns reusable mappings between entire spatial fields from many simulated examples, allowing the trained neural operators to be applied to previously unseen patterns: a forward neural operator predicts the sea-surface height (SSH) fingerprint produced by a Greenland ice-thickness change, while an inverse operator estimates the ice change from its fingerprint. To keep the inversion physically consistent, we require the inferred ice change to reproduce the input SSH fingerprint when passed through the trained forward model.
\subsection{Model Architecture}
\begin{figure*}
  \centerline{\includegraphics[width=\linewidth]{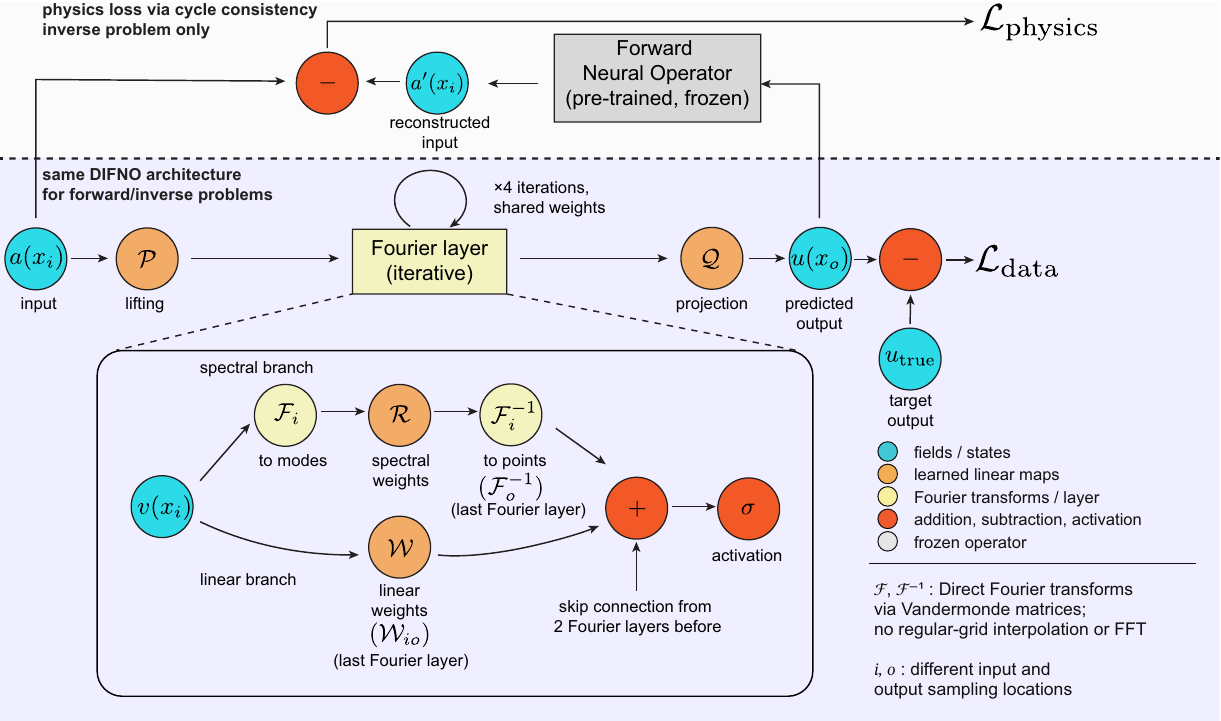}}
  \caption{Architecture of the Physics-Informed Direct Iterative Fourier Neural Operator (PI-DIFNO).  Lower block shows the DIFNO portion shared between the forward and inverse problem: the input field is lifted, passed through one Fourier layer applied four times with shared weights, and projected to the predicted output, which enters the data loss. Inside the layer, the spectral branch evaluates direct Fourier transforms on the point cloud through Vandermonde matrices; in the last Fourier layer this branch and the parallel linear branch map from the input locations to different output locations, so one operator connects mismatched domains (ice and ocean). Upper block is used for the inverse problem only: a pre-trained, frozen forward operator maps the prediction back to the input domain, and its mismatch with the original input forms a cycle-consistency physics loss. Subscripts $i$ and $o$ denote the input and output point sets, which here do not coincide. Fields (blue circles): $a(x_i)$ the input, $v(x_i)$ the latent field, $u(x_o)$ the predicted output, $u_{\mathrm{true}}$ the target, and $a'(x_i)$ the reconstructed input. Learned linear maps (brown circles): $\mathcal{P}$ (lifting), $\mathcal{Q}$ (projection), $\mathcal{R}$ (spectral weights), $\mathcal{W}$ (pointwise linear weights; $\mathcal{W}_{io}$ in the last layer, where it changes domain). Transforms: $\mathcal{F}_i$ the direct Fourier transforms ($\mathcal{F}_i^{-1}$ the inverse) built on the input points ($\mathcal{F}_o^{-1}$ in the last layer, built on the output points); $\sigma$ the activation layer. Losses: $\mathcal{L}_{\mathrm{data}}$ compares prediction with target, $\mathcal{L}_{\mathrm{physics}}$ compares reconstructed with original input.}
\label{fig:pidifno}
\end{figure*}

Consider a mapping $G:A\rightarrow U$, where $A$ and $U$ are infinite-dimensional function spaces. A neural operator $G_\theta$ approximates this mapping from a finite collection of observed input ($a\in A$) and output ($u \in U$) pairs using parameters $\theta$ in a finite-dimensional parameter space $\Theta$ within a neural network. Fourier Neural Operator (FNO, \citeA{li2021fourier}) is one of the most widely used neural operators that learns such mappings in spectral space using the Fast Fourier Transform (FFT), but it typically requires the input and output to reside on the same regular grid. Here, the ice and ocean fields are sampled on irregular point clouds that do not overlap. We develop a Direct Iterative Fourier Neural Operator (DIFNO) for this mapping, and use the same backbone for both the forward (ice-to-SSH) emulator and the inverse (SSH-to-ice) model. The inverse model additionally includes a physics-informed cycle through the frozen forward emulator, forming the physics-informed DIFNO (PI-DIFNO; Fig. \ref{fig:pidifno}).

For the backbone, DIFNO will first lift the input data to a high dimensional latent space (here 32), which is followed by a series of Fourier layers (here 4), and then projected back to output in the physical space, similar to a vanilla FNO. In a Fourier layer, a spectral branch casts the latent input $v$ into the spectral space, applies a linear transform with spectral weights $\mathcal{R}$, truncated at a maximum mode (here 20 for both directions in 2D, or 2-dimensional space), then casts back to the original latent space; another linear branch applies a linear transform of the latent input with linear weights $W$. The results from two branches are summed together, and fed to a nonlinear activation layer $\sigma$ (here Gaussian Error Linear Unit, or GELU, \citeA{hendrycks2016gelu}). The 4 layer, 32 latent dimension and 20 x 20 modes were chosen because further increasing the number of layers, Fourier modes and latent dimensions gives similar accuracy at substantially higher computational cost, while GELU is the optimal activation function here \citeA{enyeart2024some}. 

Unlike a standard FNO, DIFNO transforms data on an irregular input point cloud $x_i$ directly into spectral space using a precomputed Vandermonde matrix, an approach shown to be efficient when the retained number of Fourier modes is modest, up to 64 \cite{lingsch2024beyond}. The inverse transform can then be evaluated at a different irregular output point cloud $x_o$, allowing the input and output locations to be completely disjoint while remaining within the same coordinate domain (Fig. \ref{fig:pidifno}). This avoids interpolation between the native point clouds and a common regular grid. Compared with other available neural operators for irregular geometries, such as geo-FNO \cite{li2023geofno}, Geometry-Informed Neural Operator \cite{li2023gino}, and Mesh-Informed Neural Operator \cite{shi2025mino}, DIFNO directly connects the disjoint input and output point clouds without introducing an intermediate regular grid or a separate learned geometric mapping. The change of sampling locations is handled directly by the Fourier transforms in the spectral branch and by the cross-domain linear mapping $\mathcal{W}_{io}$ in the final linear branch; as in the standard FNO, the spectral branch is band-limited by the retained modes (about 150~km here) and the linear branch carries the finer-scale content, which is why $\mathcal{W}_{io}$ matters most for the inverse problem (Appendix \ref{app:baselines}). We adopt $\mathcal{W}_{io}$ because it is the simplest cross-domain counterpart of the pointwise linear branch of the FNO and adds little computational cost; it is specific to the ice and ocean point sets used in training, so when a trained operator must be evaluated at new sampling locations without retraining, the DIFNO without $\mathcal{W}_{io}$ (Appendix \ref{app:baselines}) or a coordinate-based mapping in its place can be used instead.

DIFNO also shares the spectral and linear weights ($\mathcal{R}$ and $\mathcal{W}$) among the Fourier layers, except for a separate $\mathcal{W}_{io}$ in the final layer where the input and output point sets differ. Weight sharing reduces the number of trainable spectral parameters and helps stabilize the training against overfitting \cite{you2022learning}. Skip connections are added across every two Fourier-layer applications to provide an additional information pathway through the iterative spectral blocks (Fig. \ref{fig:pidifno}).

The forward DIFNO is trained by minimizing the data misfit, expressed as an $L_1$ loss,
$\mathcal{L}_\mathrm{data}=|u(x_o)-u_\mathrm{true}(x_o)|$
(averaged over the training data), where $u(x_o)$ is the predicted output and $u_\mathrm{true}(x_o)$ is the ground truth. For the inverse problem, a physics-informed loss is further introduced by feeding the inverted ice mass change $u(x_o)$ into a well-trained and frozen forward DIFNO. The forward DIFNO then generates a reconstructed input $a'(x_i)$, i.e., the SSH fingerprint produced by the inverted ice field. We therefore define a cycle-consistency loss \cite{zhu2017unpaired}, $\mathcal{L}_\mathrm{physics}=|a(x_i)-a'(x_i)|$ as the physics loss. Incorporating this loss constrains the inverted ice mass change to reproduce the input SSH fingerprint through the forward ice mass change to sea level mapping, instead of producing an ice field that may be close to the ground truth in data space but that may substantially violate the governing physical relationship.

The cycle-consistency loss differs from conventional residual-based physics losses typically used in Physics-Informed Neural Networks (PINNs, \citeA{raissi2017physics}) or Physics-Informed Neural Operators (PINOs, \citeA{li2024physics}), which require explicit evaluation of the governing equations and boundary conditions. Here, the pre-trained forward DIFNO serves as a differentiable emulator of the governing sea level response, allowing the physics loss to be evaluated directly through the frozen operator without an equation-specific implementation of PDE (Partial Differential Equation) residuals or a specialized treatment of derivatives that may otherwise be needed for efficient and accurate residual evaluation \cite{wang2022respecting,ganeshram2022fc}.

\subsection{Data Generation}
\label{sec:data_generation}
We calculated the sea level fingerprints based on the theory and pseudo-spectral algorithm described in \citeA{kendall2005post}, and we solve the Sea Level Equation on a Gauss–Legendre grid truncated at spherical harmonic degree 512, using the code modified from \citeA{cederberg2023consistency}: $\Delta S=\Delta\mathcal{G}-\Delta\mathcal{R}+\Delta\Phi/g$, where $\Delta S$ is the relative sea-level change, $\Delta\mathcal{G}$ the geographically variable change of the sea-surface equipotential, $\Delta\mathcal{R}$ the radial displacement of the solid surface and $\Delta\Phi/g$ the uniform shift of the sea-surface equipotential; the SSH change is $\Delta\mathcal{G}+\Delta\Phi/g$.  The code accounts for the gravitational, rotational, and deformational response of an elastic 1-D earth to the ice/water redistribution where the depth-varying density and elastic properties are adopted from the Preliminary Reference Earth Model (PREM) \cite{dziewonski1981preliminary}. The lateral variation (a few percent) of the density and elastic parameters, as well as viscous contribution to the sea level fingerprint (a few percent) are neglected \cite{mitrovica2011robustness,coulson2022detection}. The shoreline is fixed here as the ice mass change is small enough that the shoreline migration is negligible.

By default, we consider an ice mask from \citeA{simonsen2021greenland}, and the SSH fingerprint within 55$^\circ$N to 76.3$^\circ$N, and 70$^\circ$W to 15$^\circ$W. The availability of the SSH data from satellite altimetry generally decreases towards higher latitudes, and we consider a cutoff with about 30\% or more temporal availability from 1993 to 2019 \cite{coulson2022detection}. Ocean grid points within 30 km from the shore are excluded to avoid reduced data quality \cite{climate2020coastal}, and only points in the open ocean are considered in this study to avoid complex local effects like ice-ocean interactions and fjord dynamics \cite{latuta2025drivers} (Fig. \ref{fig:setup}).

We generate three synthetic ensembles or families of Greenland ice-thickness change fields, each with 10,000 members, with progressively more complexities. All three families are built from Gaussian random fields (GRFs) sampled by truncated eigen-decomposition of a squared-exponential covariance kernel on the Greenland ice points:
\begin{equation}
C(\mathbf{x}, \mathbf{x}') = \exp\!\left(-\frac{r(\mathbf{x}, \mathbf{x}')^{2}}{2\lambda_{\mathrm{ice}}^{2}}\right),
\qquad
h(\mathbf{x}) = \sum_{j=1}^{k} \sqrt{e_{j}}\, \xi_{j}\, \varphi_{j}(\mathbf{x}),
\label{eq:grf}
\end{equation}
where $r$ is the great-circle distance between ice points $\mathbf{x}, \mathbf{x}'$, $(e_{j}, \varphi_{j})$ are the leading eigen-pairs of $C$, the truncation $k$ grows as $\lambda_{\mathrm{ice}}$ shrinks, and the coefficients $\xi_{j}$ are independent normal draws.

Starting with the first family denoted as single: each member is one GRF of Eq.~\ref{eq:grf} with a single correlation wavelength from a uniform distribution $\lambda_{\mathrm{ice}} \sim \mathcal{U}(50, 1000)$~km, and coefficients from a normal distribution $\xi_{j} \sim \mathcal{N}(m, \sigma_{\xi}^{2})$ with standard deviation $\sigma_{\xi} = 0.25$~m\,yr$^{-1}$ and mean $m \sim \mathcal{U}(-1, 1)$~m\,yr$^{-1}$, so both the spatial scale and the amplitude of the field vary across the ensemble. The second multi family superposes two such fields, $h = h^{(1)} + h^{(2)}$, with wavelengths drawn from the two ends of the range 50 to 1000~km, $\lambda^{(1)} \sim \mathcal{N}(50, 1^{2})$~km and $\lambda^{(2)} \sim \mathcal{N}(1000, 50^{2})$~km, each component carrying its own independent mean $m^{(i)} \sim \mathcal{U}(-1, 1)$~m\,yr$^{-1}$, so that fine and broad scales coexist within one sample. The single and multi families are handled with the same forward DIFNO and inverse PI-DIFNO, denoted as the default case (Fig. \ref{fig:forward}a, f).

The third coastal family is characterized by small-scale but high amplitude ice thickness change near the coast (and not in high latitudes far from the north boundary of the ocean points considered here), and low amplitude, smooth change in the interior of Greenland (Fig. \ref{fig:forward}k). This is more realistic with respect to the reference case \cite{simonsen2021greenland} compared with single and multi. It is constructed with two steps applied to one GRF $h_{0}$ from Eq.~\ref{eq:grf} with $\lambda_{\mathrm{ice}} = 50$~km and $\xi_{j} \sim \mathcal{N}(0, 0.5^{2})$:
(i) the raw field is kept only where it is negative and within $r_\mathrm{shore} \sim \mathcal{U}(100, 200)$~km from the shoreline, and south of 77$^{\circ}$N, while everywhere else it is replaced by a Gaussian-smoothed copy of itself with a smoothing width drawn from ${U}(200, 500)$~km, confining the sharp mass loss to a near-shore band;
(ii) a uniform shift sets the field mean to $m' \sim \mathcal{U}(-0.2, 0)$~m\,yr$^{-1}$.

The final ice thickness change amplitude is usually on the order of 1 m/yr (mean amplitude 1.6, 2.6, 1.4 m/yr for single, multi, coastal, respectively). The corresponding SSH fingerprint is then calculated for each of the ice fields, forming the input-output pairs for training (Fig. \ref{fig:forward}b,g,l). Before feeding the data into the neural operator, a geometric preprocessing is applied: the ice and ocean points are projected onto a polar stereographic plane (central latitude 75$^\circ$N, central longitude 45$^\circ$W, the projection used in Fig.~\ref{fig:setup}a), a common bounding box enclosing both point clouds is padded by 1\%, and each axis is rescaled independently to $[0, 2\pi]$. The Vandermonde matrices of DIFNO are precomputed on these scaled coordinates. This coordinate transform explicitly accounts for the convergence of meridians at high latitudes; the residual scale distortion of the conformal projection
across the domain is only a few percent, so a spectral mode corresponds to a nearly uniform physical wavelength in $x$ and $y$ directions, and the extent of the domain is about $3000$ km in both directions. For example, a 1-D mode with wavenumber $k^{\mathrm{sp}}=6$ corresponds to a wavelength of about 500~km.

For all models, each dataset is split at random over samples into 80\% for training and 20\% for testing (fixed seed), so training and test fields are independent realizations on the same ice and ocean point sets. SSH rates are standardized by a single mean and standard deviation computed over the full ensemble before the split (recomputing them on the training data alone changes both by less than 0.4\% of the standard deviation so there is essentially no information leaking), whereas ice-thickness changes are used in their physical units (m yr$^{-1}$); no per-sample normalization is applied, so the amplitude of each fingerprint is preserved. To characterize how information about the spatial scale of ice change is represented within the SSH fingerprints, we also decompose the centered fingerprint ensemble using principal component analysis (PCA), which provides a data-adapted orthogonal basis for the spatial variability. We compare this representation with a standard two-dimensional spectral decomposition, and later use the same fingerprint PC basis to examine the observed 1993--2019 SSH-rate field.

\subsection{Model Training}
The model training follows a two-stage procedure: we first train a forward DIFNO to predict geophysical responses to ice thickness changes, and then use the pretrained forward model to constrain the training of an inverse PI-DIFNO through cycle consistency. Although the fixed-shoreline elastic sea level response is linear and could be tabulated for a fixed set of points, the forward DIFNO provides a differentiable, batched GPU emulator that is evaluated in well under a millisecond inside the inverse training loop, and the same backbone extends without change to settings where the forward mapping is nonlinear or the sampling changes. We consider both the default and coastal configurations, with different training datasets and optimization strategies. All models are trained on NVIDIA A100 and H200 GPUs, and the training details are summarized in Table \ref{tab:models}.

As described in the model architecture, all neural operators here share the same DIFNO backbone, with $20\times20$ Fourier modes, and the width of the lifted latent dimension is 32, resulting in about 22 million trainable parameters in total. Spectral weights are initialized as in the original FNO implementation \cite{li2021fourier}, with real and imaginary parts drawn uniformly from $[0,1)$ and scaled by $1/(C_{\mathrm{in}}C_{\mathrm{out}})$, where $C_{\mathrm{in}}$ and $C_{\mathrm{out}}$ are the input and output channel numbers of the Fourier layer (both equal to the latent width, 32), and all linear layers use the PyTorch default initialization, uniform in $[-1/\sqrt{n_\mathrm{in}}, 1/\sqrt{n_\mathrm{in}}]$ with $n_\mathrm{in}$ the input dimension of the layer. In the default case, a forward DIFNO is first trained with both single and multi family data using Adam optimizer \cite{kingma2015adam}, with an initial learning rate of $10^{-3}$, reduced tenfold every 100 epochs, for 500 epochs. Once trained, the default forward DIFNO is then embedded as a frozen module for the default inverse PI-DIFNO. The total loss for the default PI-DIFNO is defined as $\mathcal{L}=\mathcal{L}_\mathrm{data}+
\lambda_\mathrm{cycle}\mathcal{L}_\mathrm{physics}$, where the weight $\lambda_\mathrm{cycle}$ is chosen so that the data loss and cycle consistency loss are balanced by their gradient norms. Similar to other multi-objective (or multi-loss) problems, we find that the gradient descent directions of the two losses often conflict with each other. We therefore adopt PCGrad (Projecting Conflicting Gradients, also known as gradient surgery, \citeA{yu2020gradient}) to handle the conflicting gradients, by projecting the gradient of $\mathcal{L}_\mathrm{physics}$ onto the normal plane of the gradient from $\mathcal{L}_\mathrm{data}$, which allows the two losses to both decrease in general during training.

When dealing with the coastal dataset, we train the coastal forward DIFNO with all three family data, since the inclusion of single and multi data helps the model to see more general random ice change patterns over the entire Greenland ice mask, and yields better performance on the coastal dataset. This yields more reliable cycle consistency loss; while for the coastal inverse PI-DIFNO, we only consider the coastal dataset alone. Since the coastal dataset is dominated by small-scale, high amplitude ice loss patches, which presents challenges for the inversion (as well as the forward problem). This is also reflected in the more severe gradient conflicts between the data and physics loss than the default case. We therefore introduce several modifications to the optimization. First, instead of a plain $L_1$ data loss, an amplitude-weighted $L_1$ data loss, $\mathcal{L}^\mathrm{weighted}_\mathrm{data}=\frac{1}{N_o}\sum_j w_j\,|u_j-u_{\mathrm{true}_j}|$ with $w_j=100\max\!\left(1,\,2|u_{\mathrm{true}_j}|/\max_{j'}|u_{\mathrm{true}_{j'}}|\right)$, where $j$ indexes the $N_o$ ice points and the maximum is taken within each sample, is used to focus on the higher true ice thickness change amplitudes in the coastal dataset than in the default one. Second,
the coastal inverse PI-DIFNO is trained with the data loss only for the first 100 epochs to warm up. Then the cycle consistency loss is introduced, with $\lambda_\mathrm{cycle}$ not only based on the ratio of gradient norms of the two losses (the data and physics misfit terms), but also scaled by epochs. The scaling linearly increases from $1/400$ to 1 from epoch 101 to epoch 500. Third, instead of PCGrad, we use CAGrad (Conflict-Averse Gradient Descent, \citeA{liu2021conflict}, convergence rate parameter $c=0.2$), which offers more control of the gradient balancing by leveraging the worst local improvement of conflicting losses to regularize the optimization trajectory.

We also tested $L_2$ loss, an area-weighted loss accounting for the latitude-dependent area represented by each point, and a loss weighted by each point's distance from the shoreline. These alternatives yield generally similar performance (the distance-to-shore weighting gives slightly worse results). Once trained, one inference (forward or inverse DIFNO) takes 0.3 ms on the H200 GPU and 2 ms on a laptop (40-core Apple M4MAX GPU), which is at least three orders of magnitude faster than solving the sea level equation on the CPU (2.5 s).

\section{Results}
The relative data errors for all forward and inverse neural operators, together with the relative cycle-consistency errors for the inverse models, are summarized in Table \ref{tab:models}.

\begin{table*}

\caption{Forward and Inverse Neural Operators in this study$^{a}$}
\label{tab:models}
\centering
\footnotesize
\setlength{\tabcolsep}{4pt}
\begin{tabular}{l l l c c}
\hline
 Model & Data & Loss & Data Error (\%) & Cycle-consistency Error  (\%) \\
\hline
 Forward, default & S+M   & $\mathcal{L}_\mathrm{data}$                     & 0.24 (0.23 / 0.25)$^{b}$        & --   \\
 Forward, coastal  & S+M+C & $\mathcal{L}_\mathrm{data}$                     & 0.35 (0.33 / 0.43 / 0.29)$^{b}$ & --   \\
 Inverse, default & S+M   & $\mathcal{L}_\mathrm{data} \& \mathcal{L}_\mathrm{physics}$$^{c}$       & 30.1 (21.8 / 38.3)$^{b}$        & 0.68 (0.69 / 0.68)$^{b}$\\
 Inverse, coastal & C     & $\mathcal{L}^\mathrm{weighted}_\mathrm{data} \& \mathcal{L}_\mathrm{physics}$$^{d}$ & 33.0                            & 3.05 \\
\hline
\multicolumn{5}{l}{$^{a}$S = single, M = multi, C = coastal. All four models share the DIFNO backbone,} \\
\multicolumn{5}{l}{\phantom{$^{a}$}forward maps ice to SSH, inverse SSH to ice. Both error columns are relative $L_1$.} \\
\multicolumn{5}{l}{$^{b}$Overall value first, per-family values in parentheses in the order of the Data column.} \\
\multicolumn{5}{l}{$^{c}$PCGrad.\quad $^{d}$Amplitude-weighted $L_1$; CAGrad, convergence rate parameter $c=0.2$.} \\
\end{tabular}
\end{table*}

\subsection{Forward Prediction of the SSH Fingerprint from Ice Mass Change}

\begin{figure*}
  \centerline{\includegraphics[width=\linewidth]{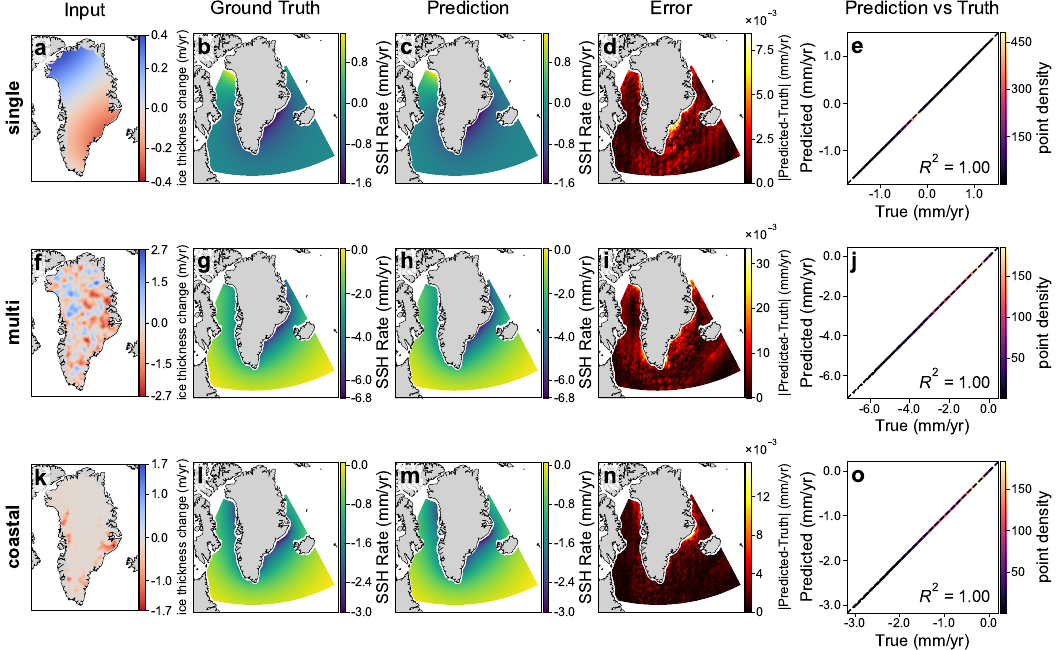}}
  \caption{Forward prediction from the ice thickness change to SSH fingerprint. Each row shows a sample from each  family of dataset: single (single-wavelength Gaussian-process ice field), multi (multi-wavelength), and coastal (ice thickness change concentrated near the coastlines); single and multi are predicted by the same default DIFNO, and coastal is handled by the coastal DIFNO. (\textit{a}) Input ice-thickness change. (\textit{b}) Ground-truth SSH rate. (\textit{c}) Predicted SSH rate. (\textit{d}) Absolute error. (\textit{e}) Prediction versus truth over the ocean points (``the one-to-one plot") with the per-sample $R^2$; black dashed line shows perfect match. (\textit{f}) to (\textit{o}) Similar to (\textit{a}) to (\textit{e}), for the multi and coastal samples.}
\label{fig:forward}
\end{figure*}

The trained forward DIFNOs accurately reproduce the numerical SSH fingerprints, with mean test errors of 0.24\% and 0.35\% for the default and coastal models, respectively (Table \ref{tab:models}). Representative samples from all three ice change families are shown in Fig. \ref{fig:forward}. The predicted and true SSH rates are nearly indistinguishable spatially, and the prediction-versus-truth points lie along the one-to-one line, with deviations of $R^2$ from unity below $2\times10^{-5}$. The accuracy also extends outside the training distributions. For the reference case of \citeA{simonsen2021greenland}, the default and coastal forward DIFNOs reproduce the pseudo-spectral solution with $R^2=0.99996$ and 0.99989 and relative errors of 0.32\% and 0.53\%, respectively. The forward approximation error is therefore much smaller than the errors of the inverse models examined below. We use the trained DIFNOs as high-fidelity emulators of the numerical sea level calculation and as differentiable forward models in the physics-constrained inversion.

\subsection{Ice Mass Change Inversion, Physical Consistency, and Non-Uniqueness}
The data errors for the inverse models are 30.1\% for the default PI-DIFNO (21.8\% for the single family and 38.3\% for the multi family) and 33.0\% for the coastal PI-DIFNO (Table \ref{tab:models}). These errors are much larger than those of the forward operators and reflect the information lost in mapping the spatially complex ice field into a smooth SSH fingerprint. The cycle-consistency errors are much smaller: less than 0.7\% for the default model and approximately 3\% for the coastal model, and 1.3\% and 3.2\% for the out-of-distribution reference case \cite{simonsen2021greenland} using the default and coastal PI-DIFNOs, respectively. Thus, the predicted ice fields reproduce the input fingerprints to within a few percent when passed through the frozen forward emulators. Training without the cycle-consistency loss produces similar ice-field data errors and visually similar predictions, while the cycle error can increase by an order of magnitude to 20\%. Without the physics loss the inversion is an unconstrained regressor whose error also spreads into directions that are observable in SSH, so its predictions no longer reproduce the input fingerprint; the physics loss plays the role of the data-misfit term of a classical inversion, confining the remaining error to the poorly resolved directions of the forward operator that underlie the non-uniqueness examined below.

\begin{figure*}
  \centerline{\includegraphics[width=\linewidth]{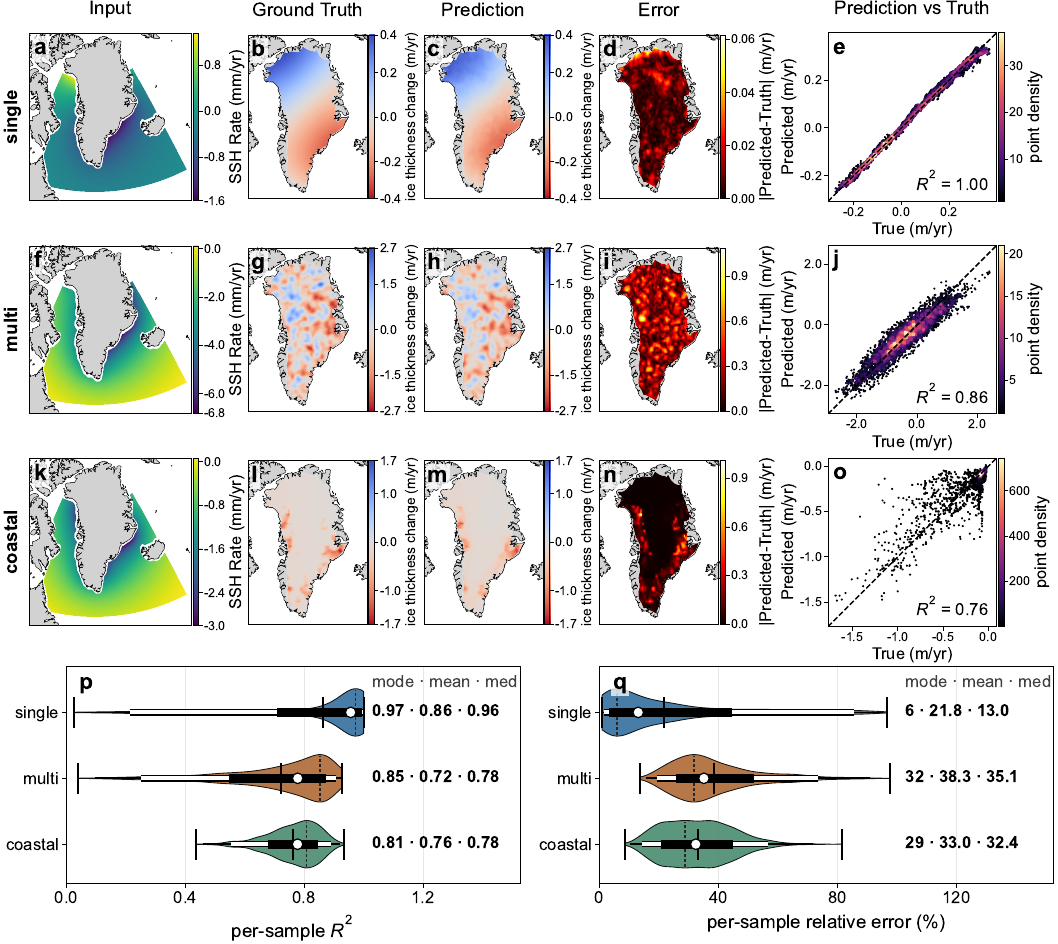}}
  \caption{Inversion from the SSH fingerprint to ice thickness change. Each row among the first three shows a sample from each family of dataset; single and multi are inverted by the same default PI-DIFNO, and coastal is handled by the coastal PI-DIFNO. (\textit{a}) Input SSH rate. (\textit{b}) Ground-truth ice-thickness change. (\textit{c}) Predicted ice-thickness change. (\textit{d}) Absolute error. (\textit{e}) Prediction versus truth over the ice points, with the per-sample $R^2$ (0.9952 here rounded to 1.00). (\textit{f}) to (\textit{o}) Similar to (\textit{a}) to (\textit{e}), for the multi and coastal samples. (\textit{p}) Violin plot with width showing the probability density distribution of per-sample $R^2$ over each family's test set. The three vertical bars are minimum, mean and maximum from left to right; alternating horizontal black and white bars cover the middle 68, 95 and 99 percentiles; the vertical dashed line shows mode; the white circle marks the median and the numbers give mode, mean, and median. (\textit{q}) Similar to (\textit{p}), for the relative error.}
\label{fig:inverse}
\end{figure*}

Representative samples for the PI-DIFNOs are also visualized in Fig. \ref{fig:inverse}. For the single family, if the ice thickness change field involves a long wavelength ($\lambda_\mathrm{ice}=1000$ km for the sample here), the PI-DIFNO can recover the full field both in terms of pattern and amplitude, with $R^2>0.995$, except for small-amplitude high-frequency noise, likely because single and multi families are trained in the same model (Fig \ref{fig:inverse}a-e). Furthermore, the distribution of $R^2$ and relative data error shows most of the samples in the single family can be well-recovered (e.g., median $R^2=0.96$, mode $R^2=0.97$), although a long tail of low $R^2$ and high error exist (Fig. \ref{fig:inverse}p, q). For the multi family, despite the mean data error of 38\% (Table \ref{tab:models}), the overall pattern and amplitude are also recovered in the presence of the 50 km ice change component, while the scatter in the one-to-one plot forms a narrow ellipse along the perfect fit dashed line(Fig \ref{fig:inverse}f-j). The inversion example from multi has $R^2=0.86$, close to the mode value of the entire family (0.85, Fig \ref{fig:inverse}p). The best inverted samples from the multi family are less resolved compared to that in the single family, while the high probability density samples cover a broader range of $R^2$ and data error (Fig \ref{fig:inverse}p, q). In the case of the coastal family, the general pattern and amplitude of the ice thickness change can be still inverted for the in-distribution test set, with slight smearing; the points in the one-to-one plot are more scattered than the multi family, but still around the perfect match dashed line (Fig \ref{fig:inverse}k-o). The coastal example is inverted with $R^2=0.76$, the same as the family mean (Fig \ref{fig:inverse}p). The overall inversion statistics are similar between multi and coastal family, but the coastal samples do not have the long bad-performance tail as in single or multi family.

\begin{figure*}
  \centerline{\includegraphics[width=\linewidth]{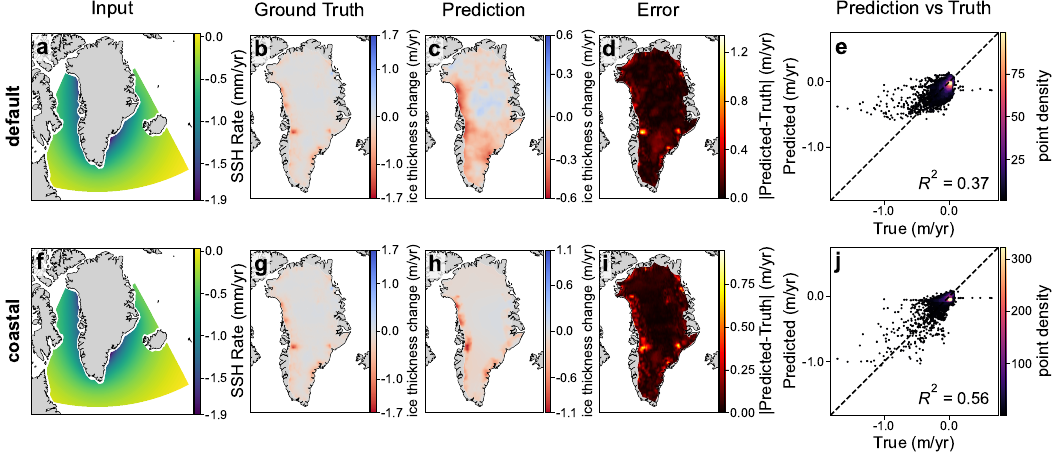}}
  \caption{Inversion of the reference case from SSH fingerprint to the satellite-altimetry-derived Greenland ice loss \cite{simonsen2021greenland}. (\textit{a}) to (\textit{e}) The default PI-DIFNO: input SSH rate, ground truth, prediction, absolute error, and prediction versus truth, as in Fig.~\ref{fig:inverse}. (\textit{f}) to (\textit{j}) Similar to (\textit{a}) to (\textit{e}), for the coastal PI-DIFNO.}
\label{fig:sshref}
\end{figure*}

We further test the two trained PI-DIFNOs on the out-of-distribution reference case from \citeA{simonsen2021greenland}. Even using the default model trained on single and multi family, the prior without preference on ice mass change locations, we can still resolve the smoothed, long wavelength near-shore ice loss pattern around the main "mascons" (mass concentrations, especially those near the west and southeast coast), although unrealistic ice accumulation starts to show up in the interior of GrIS, and the amplitude of the ice loss is only as large as 0.6 m/yr, much smaller than the ground truth (1.7 m/yr) (Fig. \ref{fig:sshref}a-d). This prediction yields $R^2=0.37$, a positive correlation, and points in the one-to-one plot are mostly near the perfect fit line, although the largest ice loss amplitudes are systematically underestimated (Fig. \ref{fig:sshref}e). However, if we use the coastal PI-DIFNO, the overall pattern and mascons are clearly better resolved, with the peak ice loss amplitude almost doubled from the default model (1.1 m/yr), and $R^2$ improved to 0.56. The ice accumulation within the Storstrømmen Glacier in the Northeast Greenland Ice Stream suggested by \citeA{simonsen2021greenland} is not resolved in the two models here, likely because the feature is relatively small and lies north of most of the ocean observations, where the SSH sensitivity is weak. (Fig. \ref{fig:sshref}). Interestingly, although the total ice mass change spans a wide range in our synthetic data (up to about $\pm 2500$~Gt\,yr$^{-1}$ for the single and multi families, and $-385$ to $+30$~Gt\,yr$^{-1}$ for the coastal family), both inverse models recover a total ice mass change close to that of the reference case ($-157$~Gt\,yr$^{-1}$ on our grid; $-160$ and $-152$~Gt\,yr$^{-1}$ from the default and coastal models, respectively), even though their $R^2$ differ substantially. The two PI-DIFNOs therefore produce ice change fields with clearly different spatial patterns and amplitudes while retaining small cycle errors for the same unseen reference fingerprint. Together with the satellite-altimetry-derived reference field \cite{simonsen2021greenland}, this gives three distinct ice change configurations that reproduce essentially the same SSH fingerprint. The reference case provides a direct example of the non-uniqueness of the inversion and of the role of the ice mass change prior in selecting among physically consistent solutions.

\subsection{An Error Law for Inversion Fidelity}

\begin{figure*}
  \centerline{\includegraphics[width=\linewidth]{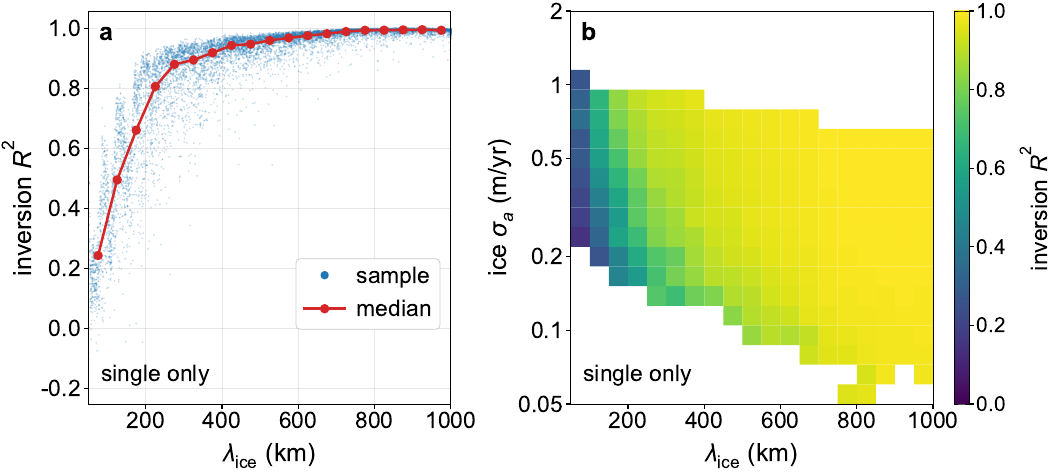}}
  \caption{Inversion $R^2$ variation with respect to the wavelength $\lambda_{\mathrm{ice}}$ and standard deviation $\sigma_a$ of the ice change field for the single family. (\textit{a}) $R^2$ versus $\lambda_{\mathrm{ice}}$, where blue dots show each sample and the red line shows the median for every 50 km bin. (\textit{b}) Heatmap showing the joint distribution of $R^2$ versus $\lambda_{\mathrm{ice}}$ and $\sigma_a$, with the median value for each bin. $\lambda_{\mathrm{ice}}$ is the correlation wavelength used to generate Gaussian random ice fields.}
\label{fig:wavelength}
\end{figure*}

The broad distributions of inversion performance in Fig. \ref{fig:inverse}p,q indicate that the resolving power of the fingerprint varies strongly among samples with different source fields. We first examine the single family, for which the source correlation wavelength $\lambda_\mathrm{ice}$ is known by construction. Figure \ref{fig:wavelength}a shows that $R^2$ remains close to 1 at long wavelengths and decreases rapidly as $\lambda_\mathrm{ice}$ approaches approximately 300~km. This behavior follows directly from the spatially smooth nature of the sea level response: the SSH sensitivity kernel decays rapidly with distance, suppressing progressively shorter-wavelength variations in the ice field. At a distance of approximately 300~km, comparable to the transition in inversion performance, the normalized sensitivity kernel has decreased to about 0.05 (Fig. \ref{fig:setup}b,c).

$R^2$ changes not only with the ice GRF wavelength $\lambda_\mathrm{ice}$, but also the ice field standard deviation $\sigma_a$. Fig. \ref{fig:wavelength}b shows the heatmap of $R^2$ in the two dimensional space of $\lambda_\mathrm{ice}$ and $\sigma_a$. While $\lambda_\mathrm{ice}$ offers the first-order control, with a fixed $\lambda_\mathrm{ice}$, as $\sigma_a$ increases, $R^2$ generally increases. For samples in the single family, for a given $\lambda_\mathrm{ice}$, $\sigma_a$ is mainly controlled by the total amplitude of the ice field (which controls the difference of maximum and minimum, or the range of the distribution), but $\sigma_a$ is also affected by the relative strength of $\xi_{j}$, the coefficient for each eigenvector of the covariance kernel (essentially the spatial pattern, or the shape of the distribution). Because the forward ice-to-SSH mapping here is linear and noise-free (for first-order elastic sea level equation with fixed shoreline, \citeA{mitrovica2011robustness}), the physical resolving power itself should not depend on a uniform rescaling of the ice-change amplitude. The observed dependence of $R^2$ on $\sigma_a$ therefore reflects the covariation of amplitude and spatial structure within the prior ensemble rather than a network-specific effect: a strictly scale-equivariant linear estimator trained on the same data shows the same behavior, and rescaling the input-target pairs leaves the DIFNO $R^2$ essentially unchanged. We separate this amplitude-dependent component empirically from the wavelength dependence below.

We start from the definition of $R^2$ and $\sigma_a$:
\begin{equation}
      R^2 = 1 - \frac{\sum_j \left( u_j - u_{\mathrm{true}_j} \right)^2}
                     {\sum_j \left( u_{\mathrm{true}_j} - \bar{u}_{\mathrm{true}} \right)^2},
      \qquad
      \sigma_a^2 = \frac{1}{N_o} \sum_j \left( u_{\mathrm{true}_j} - \bar{u}_{\mathrm{true}} \right)^2
\end{equation}
where $\bar{u}_{\mathrm{true}}$ is the spatial mean of the true ice change field for each sample, and $N_o$ is the number of output ice points. We can then define the averaged residual sum of squares, or the unexplained variance as $U$:

\begin{equation}
  U = (1-R^2)\,\sigma_a^2
    = \frac{1}{N_o} \sum_j \left( u_j - u_{\mathrm{true}_j} \right)^2.
\end{equation}

Therefore we have $R^2=1-U/\sigma_a^2$. We then consider an empirical relationship:
    \begin{equation}
    \label{eqn:law}
      U = U_0 + d_\infty\,\sigma_a^2
      \quad\Longleftrightarrow\quad
      R^2 = 1 - d_\infty - \frac{U_0}{\sigma_a^2}
    \end{equation}
where $U_0$ and $d_\infty$ are fitted constants describing two first-order contributions to the unexplained variance. In the $U$ formulation, $U_0$ is the component that does not scale with the ice-field variance, while $d_\infty\sigma_a^2$ increases in proportion to $\sigma_a^2$. The equivalent $R^2$ formulation separates their effects on inversion fidelity. The term $U_0/\sigma_a^2$ causes $R^2$ to increase as the ice change amplitude increases and becomes progressively less important at large $\sigma_a$. In the large-amplitude limit this term vanishes, and $R^2$ approaches the ceiling
$R^2_\mathrm{ceiling}=1-d_\infty$. Thus, $U_0$ controls the amplitude-dependent approach toward the ceiling, while $d_\infty$ describes the remaining deficit from perfect recovery once the amplitude is no longer limiting.

\begin{figure*}
  \centerline{\includegraphics[width=\linewidth]{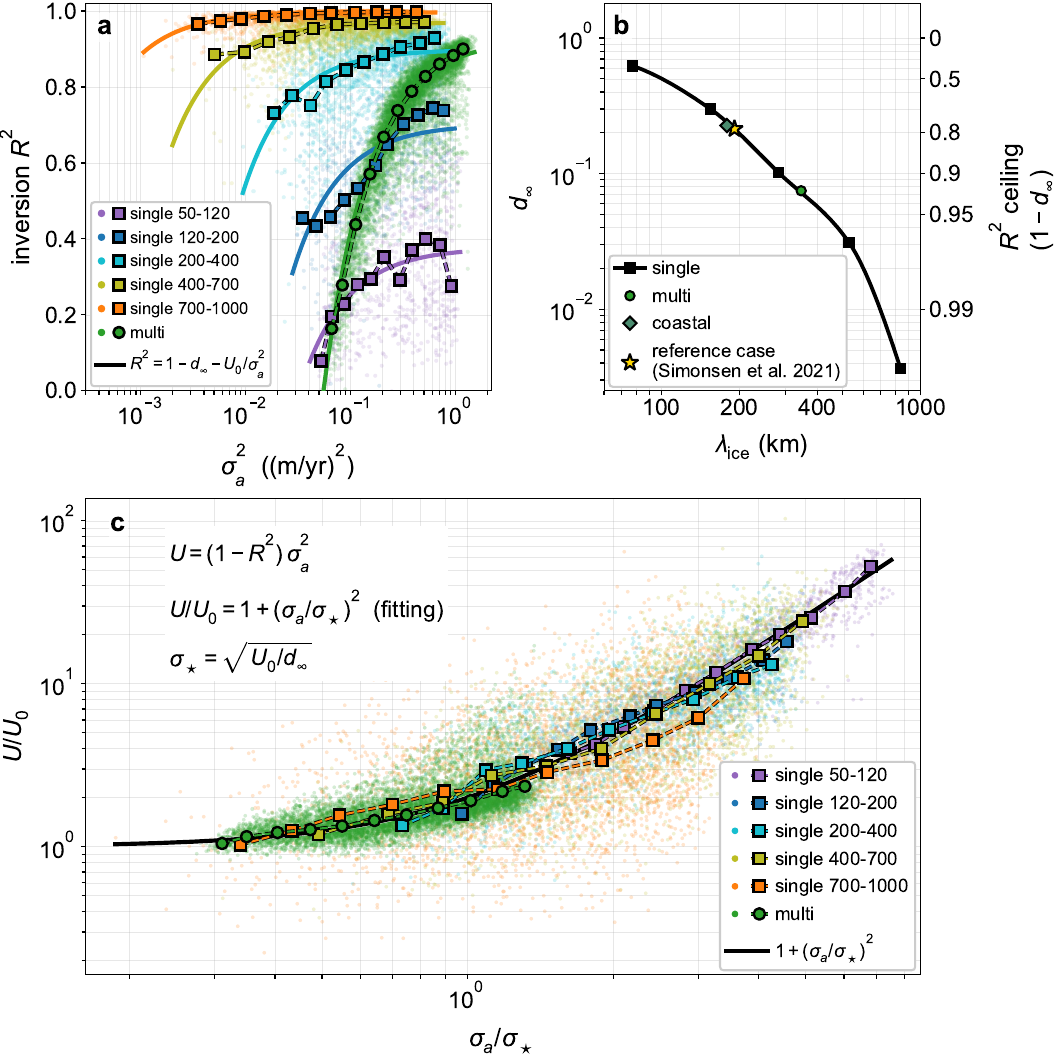}}
  \caption{The error law of the inversion. (\textit{a}) Inversion $R^2$ versus $\sigma_a^2$ for different families: single family split based on $\lambda_{\mathrm{ice}}$, and multi. Dots show each sample, dashed lines show median for each family in each $\sigma_a^2$ bin, with square markers for single and circles for multi. Solid lines are best-fitting based on the error law $R^2 = 1 - d_\infty - U_0/\sigma_a^2$. (\textit{b}) $d_\infty$ versus $\lambda_{\mathrm{ice}}$. Black line shows the fitted $d_\infty$ for single family at different $\lambda_{\mathrm{ice}}$ as in (\textit{a}), the circle and diamond show the error-law fitted $d_\infty$, and the equivalent single correlation wavelength for multi and coastal family, respectively. The star shows the predicted $d_\infty$ based on the black line and equivalent single correlation wavelength for the reference case from \citeA{simonsen2021greenland}. (\textit{c}) With $U = (1-R^2)\,\sigma_a^2$, single and multi families fall onto the error law parabola $U/U_0 = 1 + (\sigma_a/\sigma_\star)^2$, with $\sigma_\star = \sqrt{U_0/d_\infty}$. Legends are similar to (\textit{a}), with the solid line showing the error law.

  }
\label{fig:errorlaw}
\end{figure*}

Figure \ref{fig:errorlaw}a shows the dependency of $R^2$ on $\sigma_a^2$ for different wavelength bands for the single family, the dashed lines with square markers show the median $R^2$ for different $\sigma_a^2$. We calculated the best fitting curve based on Equation \ref{eqn:law} and find they can explain the first-order trend of all these median curves well (fitting shown as solid lines Fig. \ref{fig:errorlaw}a). This fitting can be extended to the multi family and it also closely matches the median curve (green solid and dashed lines in Fig. \ref{fig:errorlaw}a). While these curves occupy a different region of the $R^2$ versus $\sigma_a$ space for different wavelengths and families (since the fitted $U_0$ and $d_\infty$ vary), they can be further collapsed onto the same curve, by reformulating Equation \ref{eqn:law} as:
\begin{equation}
\label{eqn:law_final}
  \frac{U}{U_0} = 1 + \left(\frac{\sigma_a}{\sigma_\star}\right)^{2},
  \qquad \sigma_\star = \sqrt{U_0/d_\infty}.
\end{equation}
In this form, the family- and wavelength-dependent parameters are absorbed into the normalized amplitude $\sigma_a/\sigma_\star$. In this case, the samples and median curve for each subset in the reformulated space indeed collapse to the expected parabola, as shown in Fig. \ref{fig:errorlaw}c. On the left side of this parabola, $U/U_0$ approaches 1, indicating that the samples are dominated by $U_0$ (large $\sigma_\star$), which corresponds to the rapidly descending $R^2$ and small $\sigma_a^2$ (or the floor of $R^2$) in Fig. \ref{fig:errorlaw}a; on the right side of the parabola, $d_\infty$ dominates (small and  $R^2$ approaches $R^2_\mathrm{ceiling}$, which corresponds to the flattened plateau in Fig. \ref{fig:errorlaw}a. Equation \ref{eqn:law_final} can be thus used as an empirical error law to gain insight into the inversion.

Using the error law here, we can now exclude the effect of $\sigma_a$ (and therefore amplitude), and focus on $d_\infty$, which describes the $R^2_\mathrm{ceiling}$ at different $\lambda_\mathrm{ice}$ when $\sigma_a$ is no longer a limiting factor. We find for the single family, $d_\infty$ can change more than two orders of magnitude over the range of GRF wavelength $\lambda_\mathrm{ice}$ considered here (from $<4\times 10^{-3}$ near 1000 km to $>0.6$ near 50 km). As a result, $R^2_\mathrm{ceiling}$ decreases from near 1 to $<0.4$ with decreasing $\lambda_\mathrm{ice}$ (Fig. \ref{fig:errorlaw}b). 

We can further analyze the multi family results in the same framework. Since multiple GRF wavelengths are superimposed in the multi family (and GRFs of different correlation wavelengths are not orthogonal), we define an effective single GRF wavelength from the radial autocovariance (the empirical counterpart of the GRF kernel) of the family's variability. We first fit the autocovariance with the best single GRF kernel, then repeat random decompositions (using a non-negative least-squares fit) with one to three randomly drawn correlation wavelengths, keeping only those results with misfit comparable to that of the best single kernel, until 1000 decompositions are accepted. For each retained decomposition we compute the Voigt-Reuss-Hill mean \cite{hill1952elastic} of its wavelengths weighted by the fitted coefficients, i.e. the average of the arithmetic and harmonic weighted means, and take the median over all such decompositions. This raw estimate is then calibrated against the single family, whose GRF wavelength is known. The final effective $\lambda_\mathrm{ice}$ is robust with repeated tests (standard deviation less than 5 km), and we find it is 346 km for the multi family on average; combined with the $d_\infty=0.07$ and $R^2_\mathrm{ceiling}=0.93$, the multi family as a whole closely follows the $d_\infty$ and $\lambda_\mathrm{ice}$ trend from the single family (Fig. \ref{fig:errorlaw}b). Interestingly, when we project the result from the coastal family to Fig. \ref{fig:errorlaw}b, the effective $\lambda_\mathrm{ice}=179$ km and $d_\infty=0.23$, with $R^2_\mathrm{ceiling}=0.77$, which also lies close to the single family curve (the coastal family also follows the error law, not shown in Fig. \ref{fig:errorlaw}a, c for visual clarity). Therefore, despite different families, models and priors, they all follow the wavelength dependency curve for the single family (the black curve in Fig. \ref{fig:errorlaw}b). Assuming that the $d_\infty$--$\lambda_\mathrm{ice}$ relationship extends to the reference case of \citeA{simonsen2021greenland}, its effective wavelength of 191~km gives a predicted $d_\infty=0.21$ and an $R^2$ ceiling of approximately 0.79 (Fig. \ref{fig:errorlaw}b). This value represents an empirical matched-prior ceiling for the present SSH observation geometry and the GRF-based ice change distributions used to define the relationship. The reference-case $R^2$ values of 0.37 and 0.56 from the default and coastal inversions are lower because the reference field lies outside both training distributions, with the coastal prior providing a closer representation of the associated ice loss geometry. The effective $\lambda_\mathrm{ice}$ can also be viewed as a proxy for the multiscale complexity the inversion is asked to recover; reducing the fine-scale content of the target field increases the attainable $R^2$.

\subsection{Where Fine-Scale Ice Change Information Resides in the Smooth SSH Fingerprint}

The inversions above recover ice change structures at scales finer than the visually smooth SSH fingerprints, including samples from the single and multi families that are handled by the same PI-DIFNO. This raises two related questions: which components of the fingerprint contain information about fine-scale ice change, and what information allows fingerprints generated by the two ice change families to be distinguished? We address these questions using the PCA and spectral representations introduced in Section \ref{sec:data_generation}.

\begin{figure*}
  \centerline{\includegraphics[width=\linewidth]{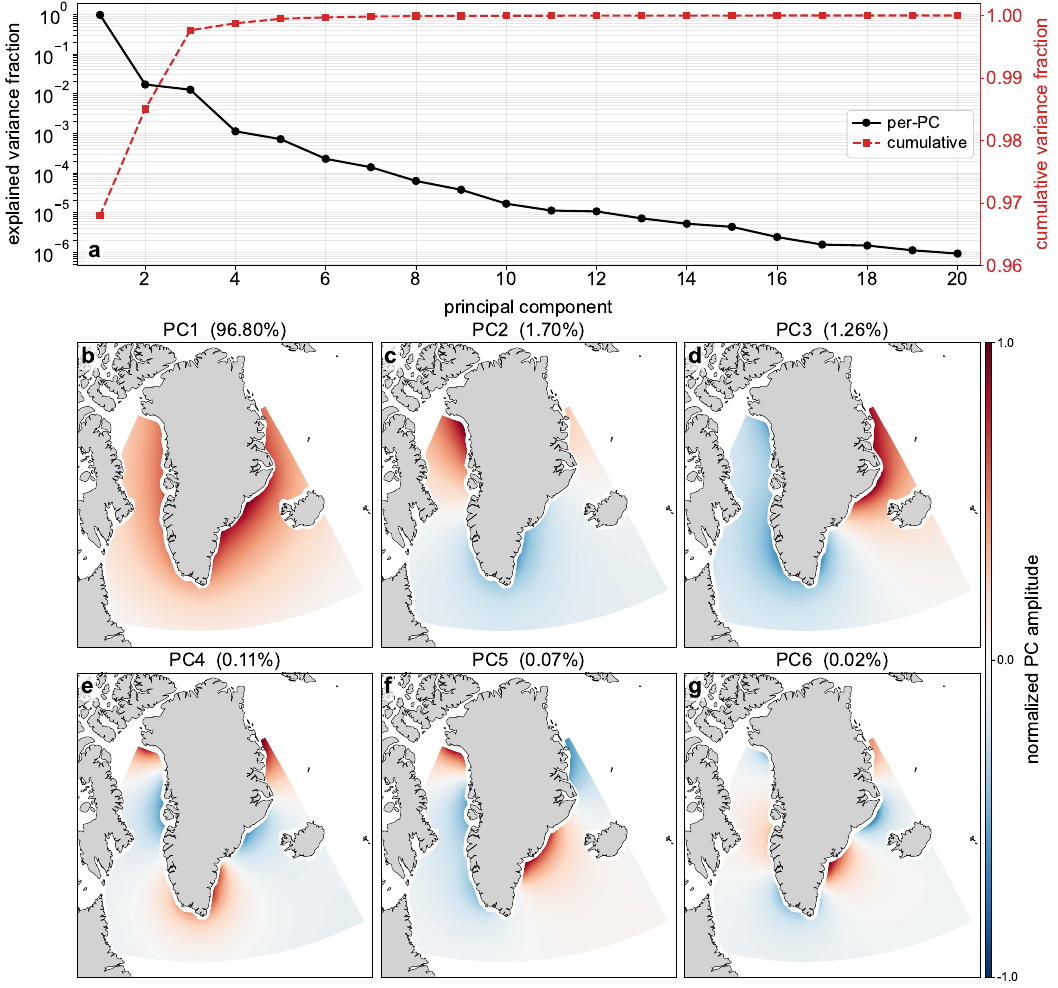}}
  \caption{Principal components (PCs) of the SSH fingerprint ensemble. (\textit{a}) Per-component and cumulative explained-variance fractions. (\textit{b}) to (\textit{g}) The spatial pattern of normalized PC 1-6, with the explained variance fraction in the title of each panel. Single and multi datasets are used but similar results are obtained using the coastal dataset.}
\label{fig:pca}
\end{figure*}

The PCA results are similar using different families and we focus on the results from the single and multi family (and the default PI-DIFNO model). The explained variance of each PC is shown in Fig. \ref{fig:pca}a: PC1 covers 96.8\% and the three leading PCs cover 99.76\% in total. Thus, only a few modes are required to represent nearly all of the fingerprint variance. The spatial pattern of the first few PCs are also shown in Fig. \ref{fig:pca}b-g. In general, PC1 monotonically decreases away from the shore and does not change sign, which essentially describes the first-order effect expected from the radially-decaying SSH sensitivity kernel (Fig. \ref{fig:setup}b). PC2 and PC3 are dipoles: each splits the ocean into two lobes of opposite sign, oriented along different directions around Greenland. From PC4 onward the patterns break into progressively more and smaller alternating-sign lobes, so each higher-order PC encodes structure at a finer scale. However, unlike spectral bases that are uniformly distributed oscillations, these patterns are adapted to the data, as each PC concentrates its lobes where the fingerprints actually vary, along and orthogonal to the coast.

\begin{figure*}
  \centerline{\includegraphics[width=\linewidth]{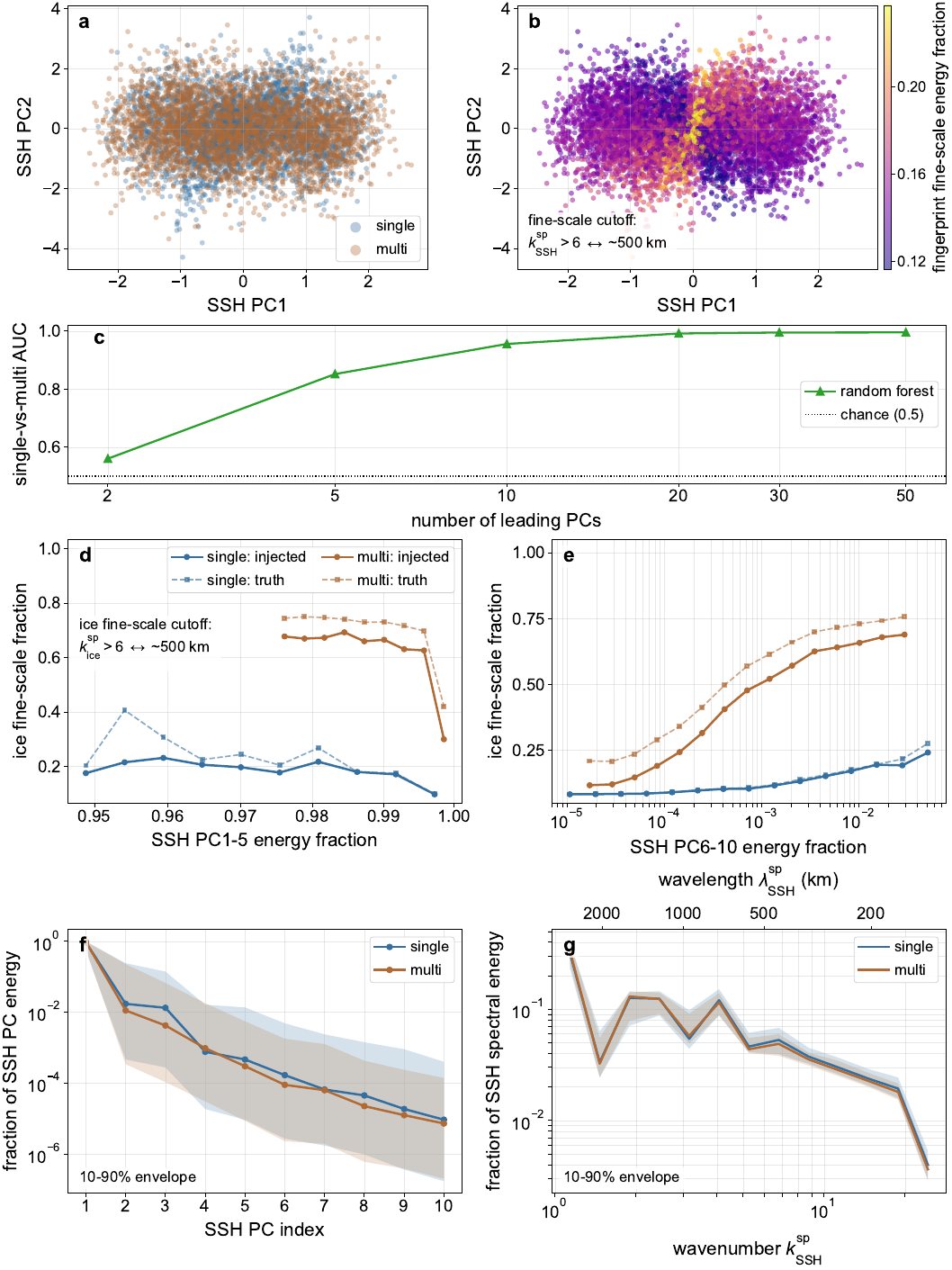}}
  \caption{The family and scale information in the SSH fingerprint. (\textit{a}) Single (blue dots) versus multi (brown dots) samples in the PC1 and PC2 space. (\textit{b}) Similar to (\textit{a}), but colored by SSH fingerprint fine-scale energy fraction, with SSH wavenumber $k_{\mathrm{SSH}}^{\mathrm{sp}} > 6$, spectral wavelengths $\lambda_{\mathrm{SSH}}^{\mathrm{sp}}<\sim 500$ km. (\textit{c}) Area under the curve (AUC) of single/multi classification performance using a Random Forest and increasing number of leading PCs, AUC of 0.5 is a random guess and 1 is a perfect classification. (\textit{d}) Ice fine-scale energy fraction (with ice wavenumber $k_{\mathrm{ice}}^{\mathrm{sp}} > 6$ and spectral wavelengths $\lambda_{\mathrm{ice}}^{\mathrm{sp}}<\sim 500$ km) versus SSH PC 1-5 energy fraction for single (blue) and multi (brown) family, ground truth (dashed line) and predicted with the default PI-DIFNO (solid line), respectively. (\textit{e}) Similar to (\textit{d}) but for SSH PC6-10. (\textit{f}) Fraction of SSH energy of each PC for single (blue) and multi (brown) family, the solid lines represent the median and the shaded region covers the 10-90$\%$ envelope. (\textit{g}) Similar to (\textit{f}) but for SSH spectral energy and each wavenumber $k_{\mathrm{SSH}}^{\mathrm{sp}}$.

  }
\label{fig:localization}
\end{figure*}

We can further visualize the fingerprints in the low-dimensional space, e.g., PC1 versus PC2 (Fig. \ref{fig:localization}a, b). Although PC1 and PC2 explain 98.5\% of the variance, we find that samples from the single and multi families are basically completely overlapped, and if we use a classifier like Random Forest \cite{breiman2001random} and try to separate the two families with only first 2 PCs, the predictive power is very close to a random guess (Fig. \ref{fig:localization}c, here we use Area Under the Curve or AUC, the curve is true positive versus false positive, AUC of 0.5 is random guess and 1 is perfect classification; the results are unchanged when each fingerprint is normalized to unit norm before projection, so the classification does not rely on the amplitude difference between families). Increasing the number of leading PCs improves the classification, and at least 20 PCs are needed to cleanly separate the ice fields with patterns emerging from different families (Fig. \ref{fig:localization}c). However, when the same PC1--PC2 space is colored by the fingerprint fine-scale energy fraction (e.g., $k_{\mathrm{SSH}}^{\mathrm{sp}}>6$, corresponding to $\lambda_{\mathrm{SSH}}^{\mathrm{sp}}\le \sim500$~km; similar results are obtained with other cutoffs), a clear angular pattern emerges across both families (Fig. \ref{fig:localization}b). Thus, PC1 and PC2 already encode the variation in the spatial scale of the SSH fingerprint, even though they contain little information for distinguishing the single and multi families; that distinction increasingly resides in higher-order PCs.

Moreover, since higher-order PCs encode smaller-scale information, we plot the ice fine-scale energy (also the ice wavenumber $k_{\mathrm{ice}}^{\mathrm{sp}} > 6$) against the energy fraction of PC1-5 and PC6-10 in Fig. \ref{fig:localization}d, e, respectively. PC1-5 represent relatively longer wavelength fingerprint information, and as their energy fraction increases, the ice fine-scale energy fraction in general decreases for both the ground truth data from single and multi families; the default PI-DIFNO also learns to inject less fine-scale energy into the ice field with more energy of longer wavelength within the SSH fingerprint, although sometimes slightly less than that of the ground truth (Fig. \ref{fig:localization}d). Conversely, for PC6-10, the opposite trend is observed, i.e., with more smaller wavelength PC6-10 energy fraction, the ice fine-scale energy fraction increases. This indicates that the amount of inverted ice field fine-scale information also comes from the small scales in the sea level fingerprint, here the higher-order PCs. 

The PCA basis also separates ice change-scale information more clearly than the regular spectral decomposition for the analyses presented here. Figure \ref{fig:localization}f shows that the fraction of SSH energy carried by individual PCs beyond PC1 varies by more than two orders of magnitude among samples, even within the 10--90\% range. By contrast, the energy fractions of the corresponding SSH spectral modes occupy a much narrower range (Fig. \ref{fig:localization}g), consistent with the similar overall smoothness of the fingerprints in spectral space. The SSH fingerprints are therefore strongly low dimensional in total variance, while weak higher-order PCs retain information needed to distinguish the fine-scale structure of the ice change fields.

\subsection{Principal Component SSH Fingerprint Projection of the Observed SSH Rate}
The synthetic experiments above use pure sea level fingerprints, whereas the observed SSH rate contains sterodynamic and other sea level contributions in addition to the GrIS fingerprint. We therefore use the PCA basis to recover only the dominant fingerprint geometry from the observed 1993--2019 SSH-rate field of \citeA{coulson2022detection}. The leading PCs are well suited to this purpose because PC1 accounts for 96.8\% and PC1--3 account for 99.76\% of the synthetic fingerprint variance (Fig. \ref{fig:pca}), with characteristic spatial scales greater than approximately 1000~km. The adopted SSH-rate field has already been corrected for the dominant wind-coherent ocean-dynamic variability following \citeA{coulson2022detection}.

\begin{figure*}
  \centerline{\includegraphics[width=\linewidth]{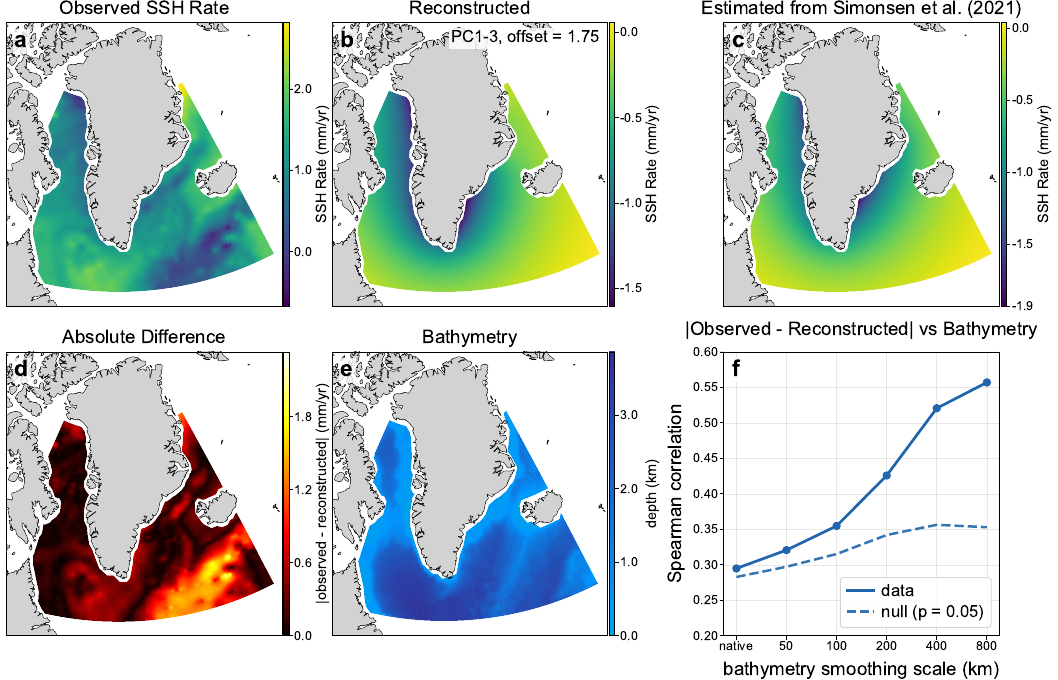}}
  \caption{Principal component analysis applied to the observed SSH Rate. (\textit{a}) Observed SSH rate from \citeA{coulson2022detection}. (\textit{b}) Its reconstruction from the leading three SSH fingerprint PCs plus a uniform offset, by fitting the gradient of SSH Rate. (\textit{c}) The reference case from \citeA{simonsen2021greenland}. (\textit{d}) The residual, $|\mathrm{observed}-\mathrm{reconstructed}|$, i.e., absolute difference between (\textit{a}) and (\textit{b}). (\textit{e}) Bathymetry. (\textit{f}) Spearman correlation between the residual in (\textit{d}) and the bathymetry in (\textit{e}), with different spatial scales of smoothing of the bathymetry (solid line), where native means no smoothing; the dashed line is the $p=0.05$ threshold from 10,000 phase-randomized residue as the null baseline.}
\label{fig:observed}
\end{figure*}

The observed mean SSH rate within our ocean mask (Fig. \ref{fig:observed}a) is based on reprocessed altimetry records \cite{taburet2019duacs}, gap-filled at high latitudes, and corrected for ongoing glacial isostatic adjustment and the fingerprints of Greenland peripheral glacier and nearby glacier (e.g., Iceland, Svalbard and the Canadian Arctic) mass changes \cite{coulson2022detection,coulson2026inversion}. SSH changes projected onto the leading spatiotemporal modes of regional wind-stress curl have also been removed \cite{coulson2022detection}. The remaining field contains the GrIS fingerprint, residual sterodynamic variability that is expected to be strongest in the subpolar gyre south of Greenland and Iceland \cite{coulson2022detection}, and the approximately uniform far-field contributions from other meltwater sources and mean ocean thermal expansion \cite{coulson2026inversion}.

We estimate the large-scale fingerprint component by fitting the spatial gradient of PC1 to the observed SSH-rate gradient while allowing a uniform offset in the observed field. The search is performed at 0.05~mm\,yr$^{-1}$ intervals using the half of the ocean points with the smallest gradient misfit, reducing the influence of spatially localized departures from the leading fingerprint geometry. The best-fitting offset is 1.75~mm\,yr$^{-1}$, consistent with the independently estimated range of 1.73--1.78~mm\,yr$^{-1}$ from the fixed-pattern inversion of \citeA{coulson2026inversion}. We then project the shifted SSH-rate field onto the fingerprint basis and retain PC1--3. The resulting reconstruction (Fig. \ref{fig:observed}b) reproduces the broad near-field geometry of the reference fingerprint calculated from the satellite-altimetry-derived ice change field of \citeA{simonsen2021greenland} (Fig. \ref{fig:observed}c). Repeating the gradient fit using PC1--3 changes the inferred offset to 1.5~mm\,yr$^{-1}$ but does not change the main spatial pattern. We do not apply the PI-DIFNO directly to the observed SSH-rate field because the inverse model is trained on pure, noise-free fingerprints, so residual sterodynamic variability could be interpreted as fingerprint structure and mapped into spurious ice-change features. We do not invert this reconstructed field for ice mass change because only the three dominant fingerprint degrees of freedom are retained, while the preceding analysis shows that finer ice mass change structure depends on the weaker higher-order modes.

The absolute residual between the observed and reconstructed SSH-rate fields is shown in Fig. \ref{fig:observed}d. Its largest amplitudes occur in the subpolar gyre region, particularly south of Iceland and near the Reykjanes Ridge. The residual also shows a spatial association with bathymetry (Fig. \ref{fig:observed}d,e, bathymetry is expressed as depth). At the native bathymetric resolution, the Spearman correlation between residual magnitude and bathymetry is $R_S=0.30$ and is significant relative to the phase-randomized null distribution. The correlation strengthens as the bathymetry is smoothed, exceeding $R_S=0.55$ at 800~km smoothing (Fig. \ref{fig:observed}f). In comparison, the observed SSH-rate field itself has a correlation magnitude below 0.15 with bathymetry for all smoothing scales considered ($p>0.44$). The bathymetric association therefore emerges primarily in the component that remains after removal of the leading fingerprint modes and is strongest at basin scales. The residual also contains fingerprint components beyond PC3, although these account for only 0.24\% of the total fingerprint variance in the synthetic ensemble.

\section{Discussion}
\subsection{Scale-Dependent Resolving Power of Sea-Level Fingerprints}

The spatial resolution of the inversion follows directly from the smoothness inherent to the sea-level response. A localized ice change affects SSH over a broad region, while the sensitivity decreases rapidly with distance from Greenland (Fig. \ref{fig:setup}). Shorter-wavelength variations in ice mass change therefore produce progressively weaker differences in the fingerprint. Our experiments show the consequence of this trend directly: substantially different ice change fields can reproduce essentially the same SSH fingerprint while maintaining small cycle-consistency errors. The dependence on the training distribution is analogous to regularization in classical geophysical inversion, where prior information stabilizes spatial variations that are relatively weakly constrained by the observations \cite{backus1968resolving,backus1970uniqueness,parker1994geophysical}.

A closely related behavior was found by \citeA{coulson2026inversion} in direct Bayesian inversions of the observed Greenland SSH rates. When ice mass change was allowed to vary at high spatial dimension, the SSH observations could be fit by unrealistic ice change patterns; restricting the ice change to a small number of regions associated with the major drainage basins produced more stable estimates and explicit posterior uncertainties. Our experiments examine the same resolving-power problem from a complementary direction. By systematically varying the wavelength, amplitude, and spatial distribution of the ice change in noise-free synthetic experiments, we can isolate the information supplied by the fingerprint itself and quantify how the recoverable spatial detail changes with the scale of the ice mass change.

The error law provides this quantitative link. The asymptotic deficit $d_\infty$ changes by almost three orders of magnitude across the range of ice change wavelengths considered here, while the multi and coastal families approximately follow the same relationship when represented by their equivalent wavelengths. In the case of the spatial complexity of the reference field, the relationship gives an empirical fingerprint-only ceiling of $R^2\sim0.79$ under the present SSH observation geometry and a matched ice change prior. The PCA analysis provides the corresponding view in SSH space: PC1--3 explain 99.76\% of the fingerprint variance, but distinguishing finer ice change patterns requires progressively weaker higher-order PCs. Thus, the same loss of spatial resolution appears in the sensitivity kernels, the wavelength-dependent error law, and the distribution of information across the fingerprint PCs. Taken together, these diagnostics show that the fingerprint acts as a scale-selective information filter: broad ice-loss structure is robustly retained, whereas progressively finer structure becomes increasingly dependent on prior information. This is the expected behavior of an ill-conditioned smoothing operator.

\subsection{Implications for Observations and Greenland Fingerprint Attribution}

The spatial concentration of information about Greenland ice loss also has implications for the SSH observations used in fingerprint inversion. Both the sensitivity kernel and the spatial structure of the leading fingerprint modes emphasize the Greenland near field. A complementary observation-mask test confirms this concentration: a coastal PI-DIFNO retrained using only SSH observations 50--250~km from Greenland retains nearly the full in-distribution inversion skill, with only a modest reduction for the out-of-distribution reference case (Appendix \ref{app:nearshore}, Fig. \ref{fig:band}). Thus, additional SSH sampling is particularly valuable in the Greenland near field for locating ice loss, while far-field observations remain useful for constraining the broad fingerprint and separating other sea level contributions. This result is also consistent with \citeA{coulson2026inversion}, whose basin-scale estimates varied more with the shoreline exclusion distance than with the SSH temporal-coverage threshold, although the overall Greenland mass-loss estimates remained relatively stable across the tested masks. Two observational requirements emerge from these results: sustained sampling of the Greenland near field and separation of weak fingerprint structure from sterodynamic SSH variability.

This concentration places particular value on sustained and accurate near-coastal SSH observations around Greenland, where conventional satellite altimetry is also more difficult because of land contamination, sea ice, gaps at high latitude, and complex fjord and shelf environments \cite{cazenave2022sea}. Dedicated coastal reprocessing has already recovered multidecadal altimetric sea level records within 20~km of coastlines \cite{cazenave2022sea}. The Surface Water and Ocean Topography (SWOT) mission further provides two-dimensional, wide-swath SSH measurements at much finer spatial resolution than conventional nadir altimetry, resolving ocean signals at wavelengths of order 10~km and, under favorable conditions, even shorter \cite{fu2024swot}. Such observations provide a direct opportunity to better sample the near-field SSH gradients identified here. The present SWOT record is, however, too short to replace the multidecadal SSH trends required for this analysis, so continuity and cross-mission consistency remain important.

The observed SSH projection illustrates the second requirement. The leading fingerprint PCs recover a large-scale spatial pattern similar to the independently predicted GrIS fingerprint, consistent with the detection of \citeA{coulson2022detection}. The synthetic experiments indicate that information about progressively finer ice change structure is carried by weaker higher-order fingerprint modes. In the observed field, these modes occupy spatial scales at which sterodynamic variability remains substantial. Ocean eddies, fronts, boundary currents, and circulation changes generate SSH variability over mesoscale and sub-basin scales \cite{chelton2007global,xu2022spatial}. Recovering finer ice loss structure from SSH therefore requires separating progressively weaker fingerprint components from increasingly competitive ocean variability.

The bathymetric organization of the residual is consistent with this interpretation. The correlation between residual magnitude and bathymetry strengthens as bathymetry is smoothed to several-hundred-kilometer scales, indicating that the association is primarily basin scale. Steric SSH integrates temperature- and salinity-driven density anomalies over depth, while ridges, continental slopes, and basin geometry organize regional circulation and the redistribution of heat and freshwater \cite{dangendorf2021data}. In the subpolar North Atlantic, sea level and steric-height anomalies show pronounced structure around the Reykjanes Ridge, Rockall Plateau, and Greenland margins, and propagating SSH anomalies follow circulation pathways influenced by bathymetry \cite{petit2019new,chafik2019north,volkov2022interannual}. The wind-stress-curl correction of \citeA{coulson2022detection} removes a substantial wind-coherent component of this variability, while advective and propagating circulation signals can remain. The residual--bathymetry relationship is consistent with this sterodynamic contribution. Ocean reanalyses offer a complementary route to isolating the sterodynamic component by providing estimates of steric and circulation-driven SSH variability that could inform corrections to the altimetric field or constraints in a joint inversion. However, substantial inter-product differences in hydrography, heat content, and transports \cite{balmaseda2015ocean,uotila2019assessment,yao2025arctic}, together with potential dependence on assimilated altimetry, call for explicit consideration of reanalysis uncertainties.

\subsection{Prior Dependence, Probabilistic Inversion, and Broader Extensions}

The present PI-DIFNO produces one ice change field for each input fingerprint, conditioned on the ice change distribution used for training. The reference experiment shows that a deterministic inversion cannot represent the full range of ice change patterns compatible with the same fingerprint. One route toward the fingerprint-only ceiling is to improve the prior on Greenland ice change. Rather than the generic statistical families used here, a more physical prior could be constructed from ensembles of process-based ice sheet simulations, building on efforts such as ISMIP6 \cite{goelzer2020ismip6}. Present-day ice geometry and thickness \cite{morlighem2017bedmachine}, ice velocity \cite{rignot2012iceflow}, and surface mass balance \cite{noel2018racmo} could be used to initialize or constrain these simulations, producing spatial patterns and correlations that are dynamically attainable. Such a prior could reduce the mismatch seen for the reference case and move the SSH-only inversion closer to the matched-prior ceiling identified here.

A separate route is to add observations that contain information not present in the SSH fingerprint and can therefore move the inference beyond this ceiling. GRACE and GRACE-FO gravity \cite{watkins2015improved} and ice altimetry \cite{simonsen2021greenland} provide direct, complementary constraints on contemporary mass and elevation change. These observations could be assimilated jointly with SSH and the ice sheet dynamical prior in a multimodal inversion or data-assimilation framework. Assimilation of altimetry into transient Greenland ice-flow models has already demonstrated how geodetic observations can constrain quantities such as basal friction and surface mass balance \cite{larour2014assimilation}. The differentiable forward DIFNO would allow candidate ice change fields to be propagated efficiently into SSH observation space within such a framework.

A probabilistic extension would also represent the remaining non-uniqueness more directly. Deep generative representations have already been explored as structured priors for high-dimensional geophysical inversions \cite{lopezalvis2021generative}. Instead of returning one ice change field, a conditional generative model could sample an ensemble of fields whose forward fingerprints agree with the observations within uncertainty. The spread among these fields would quantify which spatial features of the ice change remain poorly constrained by SSH. Simulation-based inference provides a general framework for this problem \cite{cranmer2020frontier}, and the frozen forward emulator used here for cycle consistency could impose the same physical constraint on each sampled realization. Extending the analysis in time would additionally allow differences in temporal behavior to help separate the Greenland fingerprint from sterodynamic SSH variability \cite{coulson2022detection,chafik2019north}.

The same forward-emulator and direct-inverse strategy can also be extended from the instantaneous elastic fingerprint mapping considered here to more complex history-dependent field-to-field inverse problems, such as glacial isostatic adjustment (GIA). In GIA, the forward sea level response depends on both the ice-loading history and the rheological structure of the lithosphere and mantle \cite{kendall2005post}. Recent machine learning studies have demonstrated the feasibility of accelerating parts of this forward problem: GEORGIA predicts global relative sea level for varying ice histories under a single 1-D Earth rheology \cite{lin2023georgia}, while \citeA{love2024giasurrogate} emulate the influence of 3-D Earth structure over a selected viscosity parameter space. These studies establish the value of machine-learning emulators for GIA, while leaving the jointly variable, spatially distributed ice-history and 3-D Earth-rheology problem largely open. A natural extension of the framework developed here would combine a more general field-to-field GIA emulator, allowing long-term ice-loading histories and 3-D viscoelastic Earth structure to vary jointly and potentially incorporating dynamical ice-sheet evolution, with a direct inverse operator mapping heterogeneous observations---such as relative sea level, land motion, and gravity---to ice history and Earth rheology, with the frozen forward emulator enforcing physical consistency. Given the strong non-uniqueness of GIA inversions, a probabilistic direct inverse formulation would be particularly valuable.

\section{Conclusions}

We use physics-informed neural operators to quantify how much spatial information about GrIS mass change is retained in its SSH fingerprint when the ice loss geometry is allowed to vary. The forward DIFNO reproduces the numerical sea level calculation with sub-percent error and provides a differentiable emulator for the inverse problem. Broad ice mass change patterns can be recovered, while distinct fields of ice mass change can reproduce essentially the same fingerprint with small cycle errors. Inversion fidelity is controlled primarily by the spatial scale of the ice change and secondarily by its amplitude, and these dependencies can be described by an empirical error law with a scale-dependent accuracy ceiling. For the spatial complexity of the satellite-altimetry-derived reference field, the relationship gives an empirical fingerprint-only ceiling for the present inversion class of approximately $R^2=0.79$ under the present SSH observation geometry and matched-prior assumption. This provides a quantitative measure of both the spatial information retained by the fingerprint across scales and of where the inversion becomes increasingly dependent on external constraints. Independent ice sheet observations, gravity measurements, and dynamical ice sheet constraints can add information that is weakly represented in SSH and therefore allow the full ice-mass-change inference to exceed this fingerprint-only limit.

The fingerprint is low dimensional in total variance, with PC1--3 accounting for 99.76\% of the SSH variability, while weaker higher-order modes carry information required to distinguish finer-scale ice change structure. Information useful for locating ice loss is also concentrated in the Greenland near field, and retaining only about one third of the SSH locations closest to the ice sheet preserves nearly the full synthetic inversion skill. In the observed 1993--2019 SSH rate, the leading fingerprint modes recover the expected large-scale Greenland fingerprint geometry, while the remaining signal shows a basin-scale association with bathymetry consistent with unresolved sterodynamic variability. Improved near-field observations, physical ice mass change priors, and probabilistic spatiotemporal inference can therefore extend fingerprint-based source attribution beyond the limits identified here. More generally, the same combination of numerical-model emulation and physics-constrained inversion can be applied to field-to-field inverse problems in which expensive forward models link spatially distributed physical parameters to heterogeneous observations, including GIA with 3-D viscoelastic Earth structure and long-term ice-loading histories.

\appendix
\section{Additional Neural-Operator Tests}
\label{app:baselines}
\subsection{Architecture Ablation, Computational Cost, and Baseline Comparison}
Because most trainable parameters in DIFNO reside in the final cross-domain linear mapping $\mathcal{W}_{io}$, we test whether the inversion performance mainly reflects this large dense mapping or the Fourier-operator architecture. We compare DIFNO with DeepONet \cite{lu2021deeponet}, a widely used neural-operator baseline, and with a fully connected network (FCN) of comparable parameter count. Here the DeepONet has a branch net that takes the SSH rates at all ocean points and a trunk net that takes the coordinates of each ice point, and the inner product of their outputs gives the ice thickness change at that point. In contrast, the FCN maps the SSH vector directly to the ice vector through fully connected layers without any coordinate input. We also separately remove $\mathcal{W}_{io}$ and skip connections as an architecture ablation.

The coastal inverse DIFNO contains approximately 22 million trainable parameters. About 96\% of these parameters are contained in the final cross-domain linear mapping matrix $\mathcal{W}_{io}$, while the shared spectral branch contains only about 0.8 million trainable parameters. The distribution of parameters does not reflect the computational cost of the two branches. $\mathcal{W}_{io}$ accounts for approximately 8\% of the floating-point operations, 6\% of the wall time per training step, and 3\% of the training memory. By contrast, the parameter-free Vandermonde transforms in the Fourier branch account for approximately 92\% of the floating-point operations, 80\% of the wall time, and about 30\% of the training memory. These transforms require dense complex matrix products of size $N_i\times n_{\mathrm{modes}}$ ($N_i$ is the number of input points and $n_{\mathrm{modes}}=20$ is number of Fourier modes) for every latent channel and repeated Fourier-layer application, with complex intermediate states retained during the backward pass. The $\mathcal{W}_{io}$ mapping is evaluated only once per forward pass with comparatively small input and output dimensions. The Fourier branch therefore remains the dominant computational component despite containing only a small fraction of the trainable parameters. The dense $\mathcal{W}_{io}$ could in principle be replaced by a sparser or coordinate-based mapping, but since its relative cost here is marginal, we only consider this form in our analysis. In terms of the skip connections, they do not change the parameter count and add negligible computational cost.

For the baseline comparison, all models are trained for the coastal inverse problem using amplitude-weighted $L_1$ data loss, without the cycle-consistency term so that the comparison isolates the data-fitting performance of the architectures. The inclusion of the cycle-consistency loss has little effect on the data error of DIFNO. We test various depths (4 to 6 layers) and widths for DeepONet and the fully connected network (FCN), also with 22 million parameters. The best DeepONet and FCN models give data errors of 54.9\% and 52.7\%, respectively, far from 33.0\% associated with the DIFNO (Table \ref{tab:baselines}). Using fewer parameters (0.5 to 22 million) in DeepONet and FCN yields similar results.

We test the contribution of $\mathcal{W}_{io}$ by removing this mapping from the coastal inverse architecture. The resulting model contains approximately 0.8 million parameters and its test data error increases from 33.0\% to 46.0\%. The ablated model nevertheless remains more accurate than substantially larger FCN and DeepONet baselines on the same task (Table \ref{tab:baselines}). Removing the skip connections increases the test data error from 33.0\% to 36.3\%, likely because the residual connections provide additional information pathways (Table \ref{tab:baselines}). Thus, both the cross-domain linear mapping and the skip connections improve the inversion, and the performance of DIFNO cannot be attributed simply to the large parameter count of $\mathcal{W}_{io}$ alone.

\begin{table}
\caption{Architecture Ablation and Baseline Comparison for the Coastal Inverse Problem$^a$}
\label{tab:baselines}
\centering
\begin{tabular}{l c c}
\hline
 Model & Params ($\times 10^6$) & Data Error (\%) \\
\hline
 DIFNO & 22.0 & 33.0 \\
 DIFNO without skip connections & 22.0 & 36.3 \\
 DIFNO without $\mathcal{W}_{io}$ & 0.8 & 46.0 \\
\hline
 DeepONet & 22.0 & 54.9 \\
 FCN & 22.0 & 52.7 \\
\hline
\multicolumn{3}{l}{$^{a}$All models trained for the same coastal inverse problem,} \\
\multicolumn{3}{l}{\phantom{$^{a}$}using amplitude-weighted $L_1$ data loss, without cycle consistency for simplicity.} \\
\multicolumn{3}{l}{\phantom{$^{a}$}For DeepONet and FCN, the best result} \\
\multicolumn{3}{l}{\phantom{$^{a}$}among different layers (4 to 6) and widths is reported.}
\end{tabular}
\end{table}

\subsection{Inversion Using Near-Shore SSH Observations}
\label{app:nearshore}
The sensitivity kernels and PCA results indicate that information about the spatial distribution of GrIS mass change is concentrated near the ice sheet. We test this directly by restricting the SSH input of the coastal inverse problem to ocean points located 50--250~km from the Greenland coastline. The restricted mask contains 1,609 SSH points, compared with 4,968 points in the full-domain coastal model. We retrain the coastal PI-DIFNO using this observation geometry while keeping the source family and training procedure otherwise unchanged.

\begin{figure*}
  \centerline{\includegraphics[width=\linewidth]{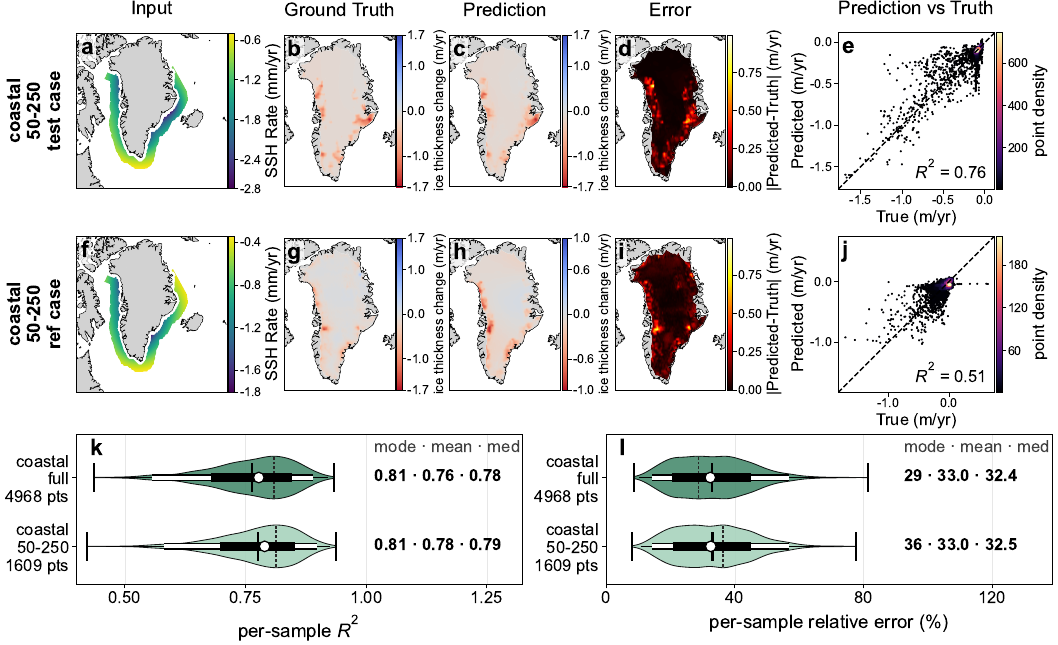}}
  \caption{Inversion from the SSH fingerprint to ice thickness change with observations restricted to a near-shore band (50 to 250 km from the Greenland coast) using a retrained coastal PI-DIFNO. The first row shows a sample from the test set. (\textit{a}) Input SSH rate. (\textit{b}) Ground-truth ice-thickness change. (\textit{c}) Predicted ice-thickness change. (\textit{d}) Absolute error. (\textit{e}) Prediction versus truth over the ice points, with the per-sample $R^2$. (\textit{f}) to (\textit{j}) Similar to (\textit{a}) to (\textit{e}), for the reference case \cite{simonsen2021greenland}. (\textit{k}) Distributions of per-sample $R^2$ for the full-domain model and the near-shore band model on the same test set, drawn as in Figure~\ref{fig:inverse}p. (\textit{l}) Similar to (\textit{k}), for the per-sample relative error.}
\label{fig:band}
\end{figure*}

The restricted model retains essentially the same performance on the held-out coastal test set (Fig. \ref{fig:band}). The full-domain model has a mean $R^2$ of 0.76 and median $R^2$ of 0.78, while the 50--250~km model gives a mean of 0.78 and median of 0.79. The mean relative data error is 33.0\% for both models, with medians of 32.4\% and 32.5\%, respectively. Thus, reducing the number of SSH observations by approximately two thirds does not reduce the inversion skill for the coastal ice change ensemble when the retained observations are concentrated in the near field.

The same restricted model is applied to the out-of-distribution reference fingerprint of \citeA{simonsen2021greenland}. The associated inversion retains the broad west- and southeast-Greenland loss pattern and gives $R^2=0.51$ (Fig. \ref{fig:band}f--j), compared with $R^2=0.56$ for the full-domain coastal model. The result is consistent with the sensitivity-kernel and PCA analyses and shows that the observations carrying most of the information for locating Greenland ice loss are concentrated within a few hundred kilometers of Greenland. However, the inclusion of broader area of SSH data helps stabilize the model and generalize to unseen scenarios out of the training data distribution.

%



%
%

\section*{Availability Statement}
Code, training data and model checkpoints for the DIFNOs can be found at \cite{bao_2026_22942700}.

\section*{Conflict of Interest declaration}
The authors declare no conflicts of interest relevant to this study.

\acknowledgments
The project was supported by the Amazon Web Services (AWS) Impact Computing Project at the Harvard Data Science Initiative (HDSI) to X.B., S.M.M. and J.X.M.. X.B. was also supported by Harvard Reginald A. Daly Postdoctoral Fellowship. The computations in this paper were partly run on the FASRC Cannon cluster supported by the FAS Division of Science Research Computing Group at Harvard University. AL acknowledges support from the Vetlesen Foundation, National Science Foundation OPP-2142592, and NASA 23-MAP23-0015.

%
%

\bibliography{agusample}

%
%
%
%
%

\end{document}